\documentclass[12pt,draftclsnofoot,onecolumn]{IEEEtran}

\usepackage{amsmath}
\usepackage{amsthm}
\usepackage{amsfonts}
\usepackage{amssymb}
\usepackage{makeidx}
\usepackage{cite}
\usepackage{graphicx}
\usepackage{mathtools}
\usepackage{cases}
\usepackage{color}
\usepackage{bbm}
\usepackage[normalem]{ulem}
\usepackage{bigints}
\usepackage{braket}
\usepackage{graphicx}

\newtheorem{theorem}{Theorem}
\newtheorem{corollary}{Corollary}
\newtheorem{lemma}{Lemma}

\newtheorem{fact}{Fact}

\newtheorem{proposition}{Proposition}
\newtheorem{remark}{Remark}

\DeclareMathOperator*{\argmin}{arg\,min}
\DeclareMathOperator*{\argmax}{arg\,max}

\DeclareMathOperator{\cV}{\mathcal{V}}
\DeclareMathOperator{\cE}{\mathcal{E}}

\DeclareMathOperator{\SINR}{\textrm{SINR}}

\DeclareMathOperator{\bR}{\mathbb{R}}

\DeclareMathOperator{\ta}{\text{(a)}}
\DeclareMathOperator{\tb}{\text{(b)}}

\DeclareMathOperator{\bP}{\mathbf{P}}
\DeclareMathOperator{\ind}{\mathbbm{1}}
\DeclareMathOperator{\bE}{\mathbf{E}}
\DeclareMathOperator{\bZ}{\mathbb{Z}}
\newcommand*\diff{\mathop{}\!\mathrm{d}}

\newcommand*\nnb{\nonumber}
\newcommand*\nnnl{\nonumber\\}

\newcommand{\uta}[1]{\underbrace{#1}_{\text{(a)}}}
\newcommand{\utb}[1]{\underbrace{#1}_{\text{(b)}}}
\newcommand{\utc}[1]{\underbrace{#1}_{\text{(c)}}}
\newcommand{\utd}[1]{\underbrace{#1}_{\text{(d)}}}

\newcommand{\ea}{\stackrel{(\text{a})}{=}}
\newcommand{\eb}{\stackrel{(\text{b})}{=}}
\newcommand{\ec}{\stackrel{(\text{c})}{=}}
\newcommand{\ed}{\stackrel{(\text{d})}{=}}
\newcommand{\ee}{\stackrel{(\text{e})}{=}}
\newcommand{\ef}{\stackrel{(\text{f})}{=}}
\newcommand{\eg}{\stackrel{(\text{g})}{=}}
\newcommand{\eh}{\stackrel{(\text{h})}{=}}

\title{An Analytical Framework for Coverage in Cellular Networks Leveraging Vehicles}
\author{Chang-Sik~Choi~and~Fran{\c{c}}ois~Baccelli
	\thanks{Chang-sik Choi is with Department of ECE, The University of Texas at Austin, TX, USA (email: chang-sik.choi@utexas.edu). Fran{\c{c}}ois baccelli is with the Department of Mathematics and the Department of ECE, The University of Texas at Austin, TX, USA (email: baccelli@math.utexas.edu)}
}
\begin{document}
	\maketitle
\begin{abstract}
This paper analyzes an emerging architecture of cellular network utilizing both planar base stations uniformly distributed in Euclidean plane and base stations located on roads. An example of this architecture is that where, in addition to conventional planar cellular base stations and users, vehicles also play the role of both base stations and users. A Poisson line process is used to model the road network and, conditionally on the lines, linear Poisson point processes are used to model the vehicles on the roads. The conventional planar base stations and users are modeled by independent planar Poisson point processes. The joint stationarity of the elements in this model allows one to use Palm calculus to investigate statistical properties of such a network. Specifically, this paper discusses two different Palm distributions, with respect to the user point processes depending on its type: planar or vehicular. We derive the distance to the nearest base station, the association of the typical users, and the coverage probability of the typical user in terms of integral formulas. Furthermore, this paper provides a comprehensive characterization of the performance of all possible cellular transmissions in this setting, namely vehicle-to-vehicle (V2V), vehicle-to-infrastructure (V2I), infrastructure-to-vehicle (I2V), and infrastructure-to-infrastructure (I2I) communications.
\end{abstract}

\section{Introduction}
The ongoing changes in the automotive industry are expected to be disruptive
in several ways. Today's vehicles have evolved from mere means of transportation
to platforms that provide multiple services, including ride sharing,
autonomous driving, data storage/processing, and Internet access.
When the next generation vehicle technology will become commercially available,
roads will be flooded with new vehicles equipped with advanced sensors,
GPS with great accuracy, high-definition cameras, as well as long and
and short-range communication devices with multiple radios.
\par Undoubtedly, a major transformation of the communication industry
and its ecosystem will follow. For instance, the IEEE WAVE standard was
established to enable vehicle-to-vehicle or vehicle-to-infrastructure
communications with a low rate \cite{karagiannis2011vehicular}.
The 3GPP has investigated vehicle to everything communications in 
the context of side link communications \cite{3gpp36211}.
New vehicular applications will require ultra fast and ultra reliable
communications for vehicles to everything networking.
The existing network architecture will be fundamentally altered,
prompting the idea of cellular networks incorporating vehicles on roads
serving as relays and/or base stations
\cite{araniti2013lte,seo2016lte,Veniam,zheng2015heterogeneous}.
\par To understand the role of vehicles in the context of future
cellular architectures, we propose a framework allowing one to
analyze the performance of a cellular network featuring vehicular
network elements in addition to the classical planar elements
(users and base stations).
The framework consists of a novel network model comprised of roads,
vehicular base stations and macro base stations.
It allows for the mathematical analysis of the 
performance experienced by users in this context.
\subsection{Related Work}
Studies of vehicular communications were mainly centered on ad hoc networking. 
A vehicle on a road was primarily identified as an apparatus to expand the
performance of ad hoc networks. The existing literature thoroughly analyzed capacity and throughput 
\cite{gupta2000capacity,grossglauser2001mobility,li2001capacity},
delay \cite{zhao2008vadd}, and routing protocols
\cite{tseng2002broadcast,perkins2003ad} in this context.
Stochastic geometry \cite{chiu2013stochastic}
provides a systematic approach to the performance analysis of large scale wireless networks
\cite{BB01,baccelli2006aloha,Baccelli2009Stochastica,haenggi2009stochastic,haenggi2009interference}.
The stationary framework of stochastic geometry allows one to investigate the performance of the typical user.
\par It was used to derive the performance of cellular networks
\cite{andrews2011tractable,blaszczyszyn2013using,keeler2013sinr}.
The Poisson network models were expanded to study multi-antenna techniques \cite{6775036,di2016stochastic} and 
new radio technologies \cite{bai2015coverage,di2015stochastic}. 
\par More recently, this tool became essential to analyze heterogeneous cellular networks that are comprised
of multiple layers of base stations distinct in their transmit powers, locations, and backhaul capabilities.
The literature on heterogeneous cellular networks first used Poisson models to study user association, coverage, and
throughput \cite{Jo2012Heterogeneous,Dhillon2012Modelinga,cheung2012throughput,heath2013modeling,Ye2013user}.
Some more recent work, including \cite{miyoshi2014cellular,chun2015modeling,li2015statistical,8089423},
proposed new spatial models for base stations and users. As discussed in \cite{Baccelli1997Stochastic},
some wireless network systems require non-Poisson spatial models particularly for vehicular networks 
since vehicles are typically on roads. The present paper can be seen as an extension of the classical Poisson cellular model to the scenario where base stations are either Poisson or Cox on Poisson lines.
\par  \cite{blaszczyszyn2009maximizing,tong2016stochastic,morlot2012population,chetlur2017coverage}
investigated the performance of wireless networks with vehicular transmitters using stochastic geometry,
by focusing on the spatial distribution of the vehicles. 
In \cite{blaszczyszyn2009maximizing,tong2016stochastic}, one dimensional spatial network models
were used to analyze the scheduling policy and reliability of communication links on roads.
Two-dimensional spatial models based on Cox point process on lines were proposed in the past
in \cite{Baccelli1997Stochastic} and recently as well to analyze the coverage probability
\cite{morlot2012population,chetlur2017coverage}. As in these 
two studies, the aim of the present paper is to analyze the coverage probability perceived by
typical users on roads. The main novelty of the present paper is the comprehensive architecture incorporating
with vehicular base stations, vehicular users, planar base stations, and planar users. To the best of our
knowledge, this framework is new and the systematic analysis of the probability of coverage of typical users of both categories which is proposed here has not been discussed so far.

\subsection{Technical Contributions}
\par \textbf{A systematic approach to characterize a heterogeneous cellular network with vehicular base stations on roads.} 
We adopt a model where the locations of vehicles are restricted to roads.
To produce a tractable model for heterogeneous cellular network with vehicles, we propose the superposition
of a stationary Cox point process on lines and a homogeneous planar Poisson point process to model two different layers of base stations,
namely vehicular base stations and planar base stations. In a similar way, a stationary planar Poisson point process
and a stationary Cox point process on lines are used to model planar users and vehicular users,
respectively. As it will become clear, the analysis will be based on a snapshot of the network  and the dynamics of vehicles is hence not  taken into account. Therefore, the Cox model proposed in this paper also captures the situation with static base stations (users) deployed along the roads because they are {topologically} equivalent to vehicular base stations (users) of our model. These two Cox point processes are built on the same stationary Poisson line process.
The proposed joint stationary framework enables a systematic approach to analyze the network performance
seen by both vehicular users and planar users. It will be evident from this paper that the type of a
user (i.e., vehicular or planar) fundamentally {alters} the statistical properties of its link,
including the distance to the nearest base station, the interference, and the coverage probability.
\par \textbf{Derivation of association, interference and coverage probability using Palm calculus.} 
To derive the performance of the proposed model, this paper considers the classical assumptions that
users are associated with their nearest base station, independently of their type, and that the received signal
power attenuates according to a path loss model with Rayleigh fading. 
Statistical properties and fundamental metrics are obtained using the joint stationarity framework and Palm calculus. 
In particular, we derive exact integral expressions for the association probability and coverage probability
under the Palm probability of each user point process. The integral expressions 
can be extended to interesting scenarios such as non-exponential fading or multi-input multi-output transceiver. 

\par \textbf{Comprehensive characterization heterogeneous wireless transmissions in cellular networks with vehicles}: We use the coverage and association expressions to inspect the performance of all wireless links in a network with both planar and vehicular transmissions. For instance, the performance of vehicle-to-vehicle (V2V) communications is captured through
	the coverage probability of the typical vehicular user by vehicular base stations, which involves
	the Palm distribution of the vehicular user point process. In the same vein, the performance of infrastructure-to-infrastructure (I2I) communications is characterized by 
	the coverage probability of the typical planar user by planar base stations. Using the Palm probability, the paper gives a comprehensive analysis of all wireless links occurring in cellular networks with vehicles, namely V2V, V2I, I2V, and I2I communications.


\section{System Model}\label{S:2}
\subsection{Spatial Modeling}
\begin{figure}
	\centering
	\includegraphics[width=0.6\linewidth]{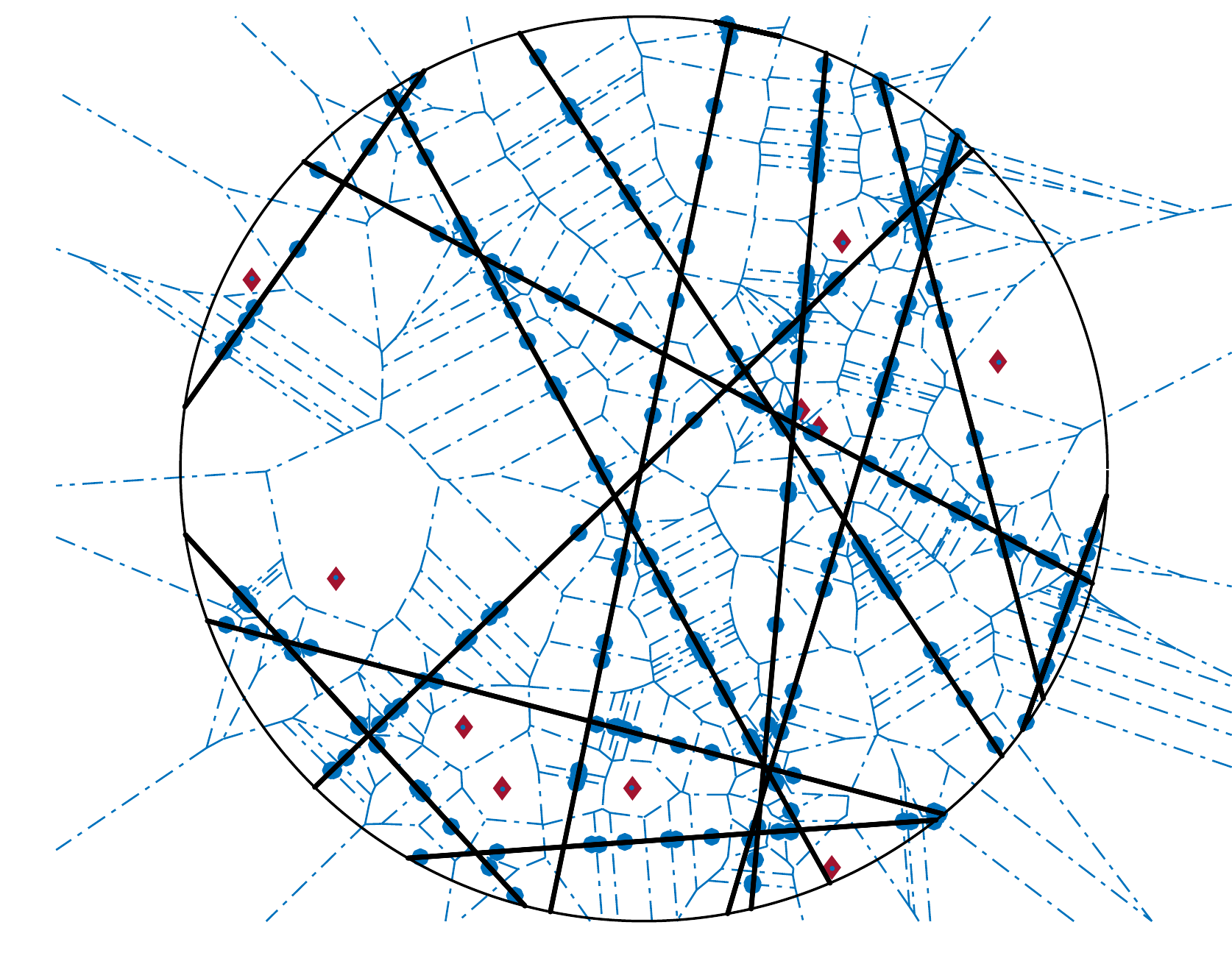}
	\caption{An illustration of roads (line), planar base stations (diamonds), and vehicular base stations (circles) with boundaries (dashed).}
	\label{fig:coxvoronoi20}
\end{figure}
\par This paper proposes a novel heterogeneous cellular network with both planar 
and vehicular base stations. Homogeneous planar Poisson point processes $ \Phi_b $, $ \Phi_u $,
with intensities $ \lambda_b $, $ \lambda_u $ are used to model macro base stations and users,
respectively. This paper calls them planar base stations and planar users respectively, to distinguish 
them from their vehicular counterparts. Both $ \Phi_b $ and $ \Phi_u $ are assumed stationary,
homogeneous, and independent. 
\par 
The vehicular base stations and users are modeled by specific Cox point processes.
To begin with, a road system is defined by an independent Poisson line process
$ \Phi_l\in\bR^2 $ produced by a homogeneous Poisson point process $ \Xi $ on the
cylinder $ \mathbf{C}:=\bR \times [0,\pi) $. More precisely, a point of $ \Xi $,
denoted by $ (r_i,\theta_i) $, describes the line $ l_i\in \bR^2 $ of equation
\begin{equation}
l(r_i,\theta_i)=\{(x,y)\in\bR^2| x \cos (\theta_i) +y \sin (\theta_i) =r_i \}, 	 
\end{equation}
where the parameters $ r $ and $ \theta $ correspond to the shortest distance from
the origin to the line and the angle between the positive $ X $ axis and 
line $ l $, respectively. For the isotropic and stationary Poisson line process,
the intensity measure of $ \Xi \in \mathbf{C} $ is given by \cite{Chiu2013Stochastica} 
\[ \Lambda_{\Xi}(\diff r \diff \theta)=\frac{\lambda_l}{\pi} \diff r \diff \theta.\]  
\par
\begin{figure}
	\centering
		\includegraphics[width=0.60\linewidth]{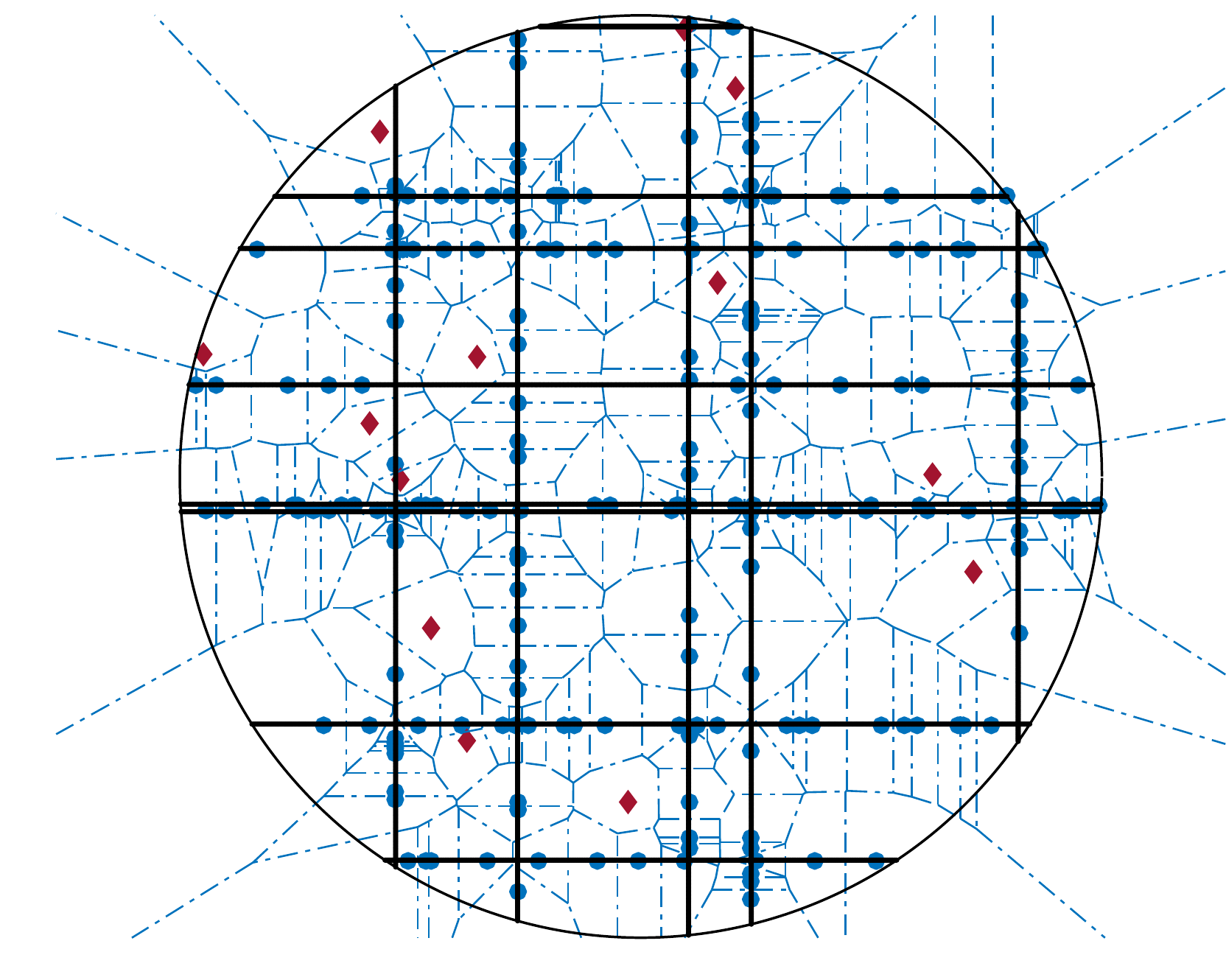}
	\caption{An illustration of roads (line), planar base stations (diamonds), and vehicular base stations (circles) with boundaries (dashed). Here the intensity over angular axis is concentrated over $ \{0,\pi/2\} $}
	\label{fig:coxvoronoi21}
\end{figure}

 Then, conditionally on the lines, vehicular base stations and users are modeled by
independent 1-D homogeneous Poisson point processes with intensity $ \mu_b $ and $ \mu_u $ {on each line},
respectively. In other words, for each undirected line produced by a point of $ \Xi $,
conditionally independent copies of Poisson processes on $ \bR $ with parameters $ \mu_b $ and $ \mu_u $
are used to describe vehicular base stations and vehicular users, respectively.
Fig. \ref{fig:coxvoronoi20} illustrates the planar base stations $ \Phi_b $ and the vehicular base stations
$ \Psi_b$ in a simulation ball of radius $ 1\text{km} $ with  $ \lambda_l=10/\text{km} $ and $ \mu_b=10/\text{km} $. 
Notice that the proposed network model is flexible in creating road systems with different topologies.
Fig. \ref{fig:coxvoronoi21} illustrates the deployment of planar and vehicular base stations when the
angular component intensity measure of $ \Xi $ is limited to two simple values:
$ \{0,\pi/2\}, $ which admits only horizontal and vertical roads.
It has a special name: a Poisson Manhattan Co'x point process. The isotropic case, where angles
are uniformly distributed, is depicted in Fig. \ref{fig:coxvoronoi20}. Remind that this paper does not study the motion of vehicles. Therefore, the vehicular base stations are topologically equivalent to static base stations deployed along the roads. We will still call them vehicular base stations in order to distinguish them from planar base stations.

\begin{table}
	\caption{Notation table}\label{T:1}
	\begin{center}
		\begin{tabular}{|l|l|}
			\hline 
			Symbols	& Description \\ 
			\hline 
			$ \Phi_b \text{ and }\Phi_u $	& Planar base stations (spatial intensity) and  planar users (spatial intensity) \\ 
			\hline 
			$ \Xi (\Phi_l)$	&  Poisson point process on cylinder set $ \mathbf{C} $ (Poisson line process on $ \bR^2 $) \\ 
			\hline
			$ l(r,\theta) $	& A line in Euclidean plane produced by a point  $ (r,\theta) \in \mathbf{C}$  \\ 
			\hline 
			$ l(r_0,\theta_0) $	& A line that contains the origin   \\ 
			\hline 
			$ \phi_l $	&  Poisson process on line $ l $ \\
			\hline 
						$ \Psi_b \text{ and }\Psi_u $	& Vehicular base stations and  vehicular users \\ 
			\hline 
			$ \bP_{\Phi_u}^0$ and $ \bP_{\Psi_u}^0 $	& Palm probabilities with respect to $ \Phi_u $ and $ \Psi_u, $ respectively \\
			\hline
			$ X_\star $ & The base station closest to the user at the origin\\\hline
		\end{tabular} 
	\end{center}
\end{table}

\subsection{Transmission Model and Performance Metrics}
This paper analyzes downlink cellular communications, namely links from base stations to users,
under the assumption of a distance-based path loss and Rayleigh fading. The received signal power
$ P_r $ at a distance $ R $ is given by $P_r=p H R^{-\alpha},$
where $ \alpha>2 $, $ p  $ is the transmit power, and $ H $ is an exponential random variable
with mean one representing the Rayleigh fading. Since both receive signal power and interference
fluctuate between users, spatially averaged performance metrics should be considered.  
In fact, all base stations and users are jointly stationary, i.e.,
$ \Phi_b,\Psi_b,\Phi_u,\Psi_u $ are $ \{\theta_t\}$-compatible on a stationary framework,
and therefore, we can use the Palm probability to study the network performance seen
by a typical user. For instance, for a given constant $ T>0 $,
the coverage probability of the typical user is defined by
under the Palm probability of  $ \Phi_u+\Psi_u $ by
\begin{equation}
p_c=\bP_{\Phi_u+\Psi_u}^0(\text{SINR}>T),\nnb
\end{equation}
where the signal to interference-plus-noise ratio (SINR) is given by the receive signal power divided
by the interference-plus-noise power and where $ \bP_{\Phi}^0 $ is the Palm probability of the point process $ \Phi $.
Similarly, the coverage probability of the typical planar (resp. vehicular) user
is simply $\bP_{\Phi_u}^0(\text{SINR}>T)$
(resp. $\bP_{\Psi_u}^0(\text{SINR}>T)$).

Assuming that noise power is negligible compared to the signal and interference powers,
the $ \SINR $ is equal to the signal to interference ratio and the coverage probability of the typical user is given by 
\begin{equation*}
p_c=\bP_{\Phi_u+\Psi_u}^0\left(\frac{pH \|X_\star\|^{-\alpha}}{\sum\limits_{X_i\in\Phi_b + \Psi_b \setminus X_\star} pH{\|X_i\|}^{-\alpha}}>T\right),
\end{equation*}
where $ X_\star $ denotes the nearest base station given by 
\begin{equation}
X_\star :=  	\argmax_{X_i\in\Phi_b + \Psi_b} \bE_{\Phi_u+\Psi_u}^0\left[\frac{pH_{X_i}}{\|X_i-0\|^{\alpha}}\right]=\argmin_{X_i\in\Phi_b + \Psi_b}\|X_i\|.\label{eq:xstar}
\end{equation} 
The expression is similar for the typical planar (resp. vehicular) user.
There are simple connections between these quantities which are discussed below.

\subsection{Preliminaries: Properties of the Proposed Cox and its Joint Typicality}\label{SS:3}
In the following, the spatial intensity of the Cox point process is evaluated.
It is classical. Yet we give its proof to better explain the proofs of later results. 
\begin{fact}\label{R:1}
	\textbf{Stationarity and spatial intensity of Cox point processes} The vehicular base stations and vehicular users have spatial intensities $ \lambda_l\mu_b $ and $ \lambda_l\mu_u  $, respectively \cite{morlot2012population}.
\end{fact}

\begin{fact}
	Consider two jointly stationary and independent point processes $ \Phi_{1}$ and $\Phi_2 . $
        Then, the Palm probability of  $ \Phi_1+\Phi_2, $ is given by 
	\begin{equation}\label{jointpalm}
	\bP_{\Phi}^0=\frac{\lambda_1}{\lambda_1+\lambda_2 }\bP_{\Phi_1}^0 	+ \frac{\lambda_2}{\lambda_1+\lambda_2 }\bP_{\Phi_2}^0, 
	\end{equation}
	where  $ \lambda_1 $ and $ \lambda_2 $ are the intensity parameters of $ \Phi_1 $ and $ \Phi_2 $, respectively.
\end{fact}
\begin{IEEEproof}
	Given in \cite{Baccelli2009Stochastica}. 
\end{IEEEproof}
We will apply the above lemma to derive the coverage probability of the typical user
\begin{align}
\bP_{\Phi_u+\Psi_u}^0(\SINR>T)
&=\frac{\lambda_u}{\lambda_u+\lambda_l\mu_u }\underbrace{\bP_{\Phi_u}^0(\SINR>T)}_{\ta} 	+ \frac{\lambda_l\mu_u}{\lambda_u+\lambda_l\mu_u }\underbrace{\bP_{\Psi_u}^0(\SINR>T)}_{\tb},\label{6}	\end{align}
where (a) is under the Palm probability of $ \Phi_u $ and (b) is under the Palm probability of $ \Psi_u $.
Therefore, in order to compute the coverage probability of the typical user, (a) and (b) need to be derived separately.
In particular, Section \ref{S:3} considers the Palm probability with respect to $ \Phi_u $ and Section \ref{S:4}
considers it with respect to $ \Psi_u. $ 

\section{Coverage: Palm Probability of $  {\Phi_u} $}\label{S:3}

In this section, we work under the Palm distribution $ \bP_{\Phi_u}^0. $ Under the latter,
there exists a typical planar user at the origin. The subscript of the Palm $ \bP_{\Phi_u}^0 $ is often omitted in the derivation.

\begin{figure}
	\centering
	\includegraphics[width=0.5\linewidth]{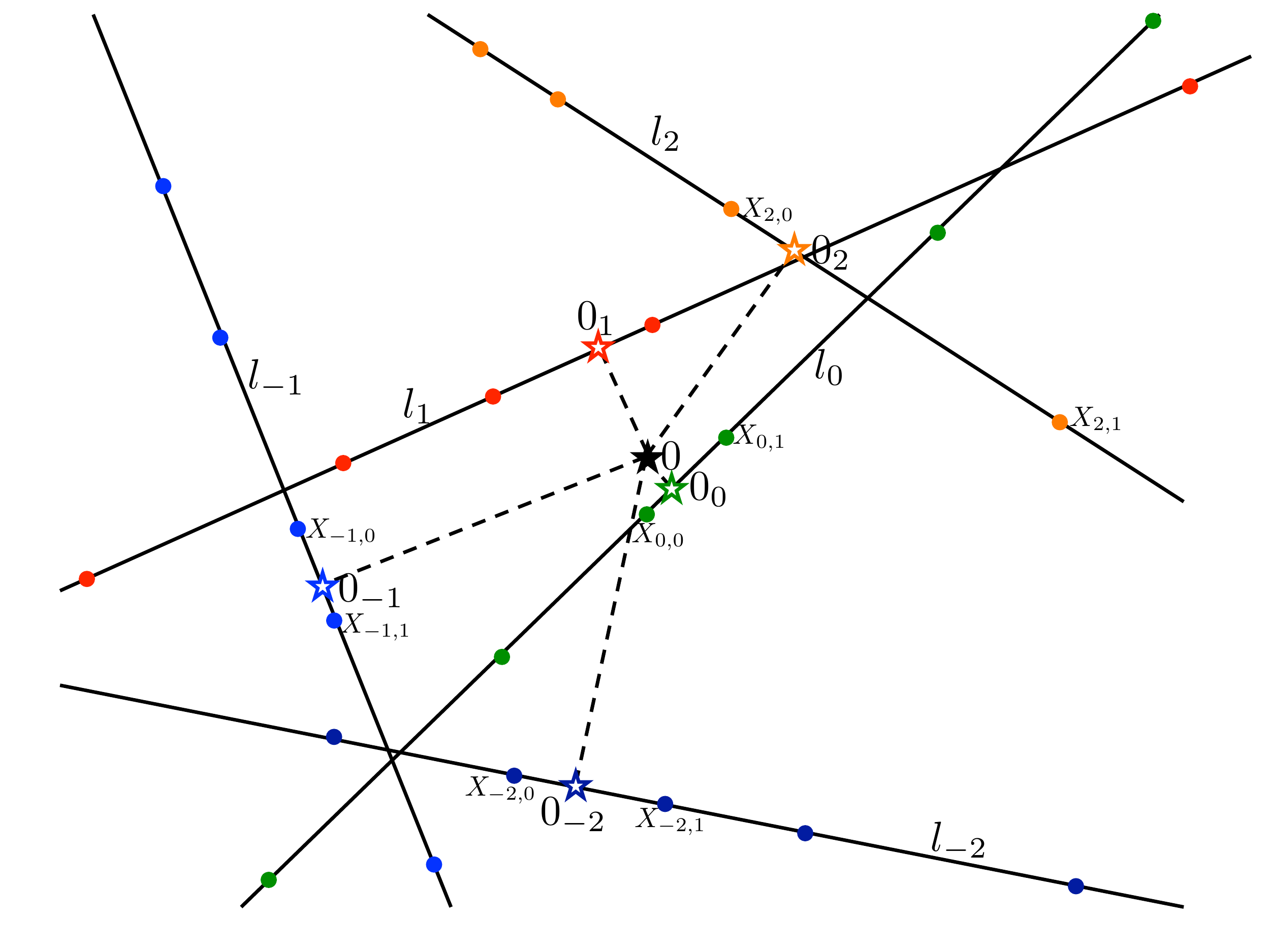}
	\caption{Illustration of the proposed numbering for the Cox point process. Notice that $ 0_{k} $ are points on the lines that are closest to the origin.}
	\label{fig:plporientation}
\end{figure}

\begin{remark}
	\textbf{Numbering of the proposed Cox points} 
	We assign two indexes $ (i,j) $ to each point; the lines are numbered by indexes $ \{i\}_{\bZ} $
according to their distances from the origin; the points of line $ i $ are numbered by indexes $ \{j\}_{\bZ} $
according to their distances from $ \{0\}_i $ which is the point of line $ i $ closest to the origin.
We utilize the classical numbering in \cite[Fig. 1.1.1]{Baccelli2013Elements}.
Fig. \ref{fig:plporientation} illustrates the proposed counting of this paper.
This numbering allows one to describe any Cox points as 
$ X_{i,j}=(t_{j(i)} \cos(\theta_i)-r_i\sin(\theta_i)\text{sgn}(\theta-\pi/2), t_{j(i)}\sin(\theta_i)+r_i\cos(\theta_i))$,
for all $ i,j\in\bZ^2 $. To visualize the points on line $ i $, consider a Poisson point process
$ \{t_{j(i)}\} $ on the $ X $ axis. These points are rotated by $ \theta_i  $
and and translated by $ \vec{v}=(r_i\sin(\theta)\text{sgn}(\theta-\pi/2) ,r_i\cos(\theta)) $, respectively.
Since an independent copy of the Poisson process is considered on each line,
we simply write $ t_{j(i)} $ by $ t_j. $ 
$
	\|X_{i,j}\|=\sqrt{t_j^2+r_i^2}.
$
\end{remark}

\subsection{Association of the Typical Planar User}
Under the Palm distribution $ \bP_{\Phi_u}^0, $ the typical user is associated with its
closest base station: either a planar base station or a vehicular base station.
We denote these events by $ X_\star\in\Phi_b $ and $ X_\star\in\Psi_b, $ respectively.
The user association is an important metric because it plays a key role to derive the coverage probability.
\par

\begin{proposition}
The probabilities that the typical planar user is associated with a planar base station and
a vehicular base station are given by 
	\begin{align}
	\bP_{\Phi_u}^0(X_\star\in\Phi_b)&=\int_{0}^{\infty}2\pi\lambda_b r e^{-\pi\lambda_b r^2-2\lambda_l\int_0^r 1-e^{-2\mu_b\sqrt{r^2-t^2}}\diff t}\diff r,\label{L4:1}\\
	\bP_{\Phi_u}^0(X_\star\in\Psi_b)&=1-\int_{0}^{\infty}2\pi\lambda_b r e^{-\pi\lambda_b r^2-2\lambda_l\int_0^r 1-e^{-2\mu_b\sqrt{r^2-t^2}}\diff t}\diff r\label{L4:2},
	\end{align}
	respectively.
\end{proposition}
\begin{IEEEproof}
	Recalling the user at the origin is a planar user, let us define two random variables
$$ R_p=\inf_{X_i\in\Phi_b}\|X_i\| \text{  and  }   R_v=\inf_{X_i\in\Psi_b}\|X_i\|. $$ 
They capture the distances from the origin to the nearest planar base station and to the nearest vehicular base station, 
respectively. By recalling the association principle in Eq. \eqref{eq:xstar}, the association probability is given by 
	\begin{align}
		\bP_{\Phi_u}^0(X_\star\in\Phi_b)&=\bE_{\Phi_u}^0\left[\ind_{R_p<R_v}\right]=\bE\left[\bE\left[\ind_{r_p<R_v}|R_p=r_p\right]\right],\nnb
	\end{align}
	where we used the independence assumption. The integrand of the above conditional expectation is given by 
	$$ 
	\ind_{R_v>r_p} \equiv \prod_{X_{i,j}\in\Psi_b} \ind_{\|X_{i,j}\|>r_p}.
	$$
	Therefore, the association probability of the typical planar user is given by 
	\begin{align}
		\bE\left[\bE\left[\ind_{R_v>r_p}|R_p\right]\right]&=\bE\left[\bE\left[\prod_{X_{i,j}\in\Psi_b}\ind_{\|X_{i,j}\|>r_p}\right]\right]\nnb\\
		&=\bE\left[\bE\left[\prod_{i,j\in\bZ^2}\ind_{r_i^2+t_{j}^2>r_p^2}\right]\right]\nnb\\
		&\ea \bE\left[\bE\left[\prod_{(r_i,\theta_i)\in\Xi}^{r_i\leq r_p}\bE\left[\prod_{t_j\in\phi_{l(r_i)}}\ind_{t_{j}^2>r_p^2-r_i^2}|\Phi_l\right]\right]\right]\nnb\\
		&\eb\bE\left[\bE\left[\prod_{(r_i,\theta_i)\in\Xi}^{r_i\leq r_p} \exp\left({-\mu_b\int 1-\ind_{t^2>r_p^2-r_i^2}}\diff t\right)\right]\right]\nnb\\
		&\ec \bE\left[e^{-2\lambda_l\int_{0}^{r_p}1-e^{-2\mu_b \sqrt{r_p^2-t^2}} \diff t}\right],\label{12'}
		\end{align}
where (a) follows from the fact that, conditionally on $ \Phi_l $, the Poisson point processes
on lines are independent. We get (b) by using the PGFL formula on the Poisson point process
$ \phi_{l(r_i,\theta_i)} $, and (c) by utilizing the PGFL formula of the Poisson point process $ \Xi $ on
the cylinder. The formula \eqref{12'} is integrated with the density function of $ R_p, $ given by
		\begin{equation*}
		f_{R_p}(w)= \partial_w \left(1-e^{-\pi\lambda_bw^2}\right)=2 \pi\lambda_bw e^{-\pi \lambda_bw^2 },
		\end{equation*}
where $f_{X}(x) $ denotes the probability distribution function of random variable $ X. $
As a result, the association probability is given by 
		\begin{align*}
				\bP^0(X_\star\in\Phi_b)&=\int_{0}^{\infty}2\pi\lambda_b w e^{-\pi\lambda_b w^2 -2\lambda_l\int_0^w 1-e^{-2\mu_b\sqrt{w^2-t^2}}\diff t}\diff w.
		\end{align*} Notice that $ \bP^0(X_\star\in\Psi_b)=1-\bP^0(X_\star\in\Phi_b)$ since $ (X_\star\in\Psi_b)={(X_\star\in\Phi_b)}^c $ .
\end{IEEEproof}


\begin{figure}
	\centering
	\includegraphics[width=0.60\linewidth]{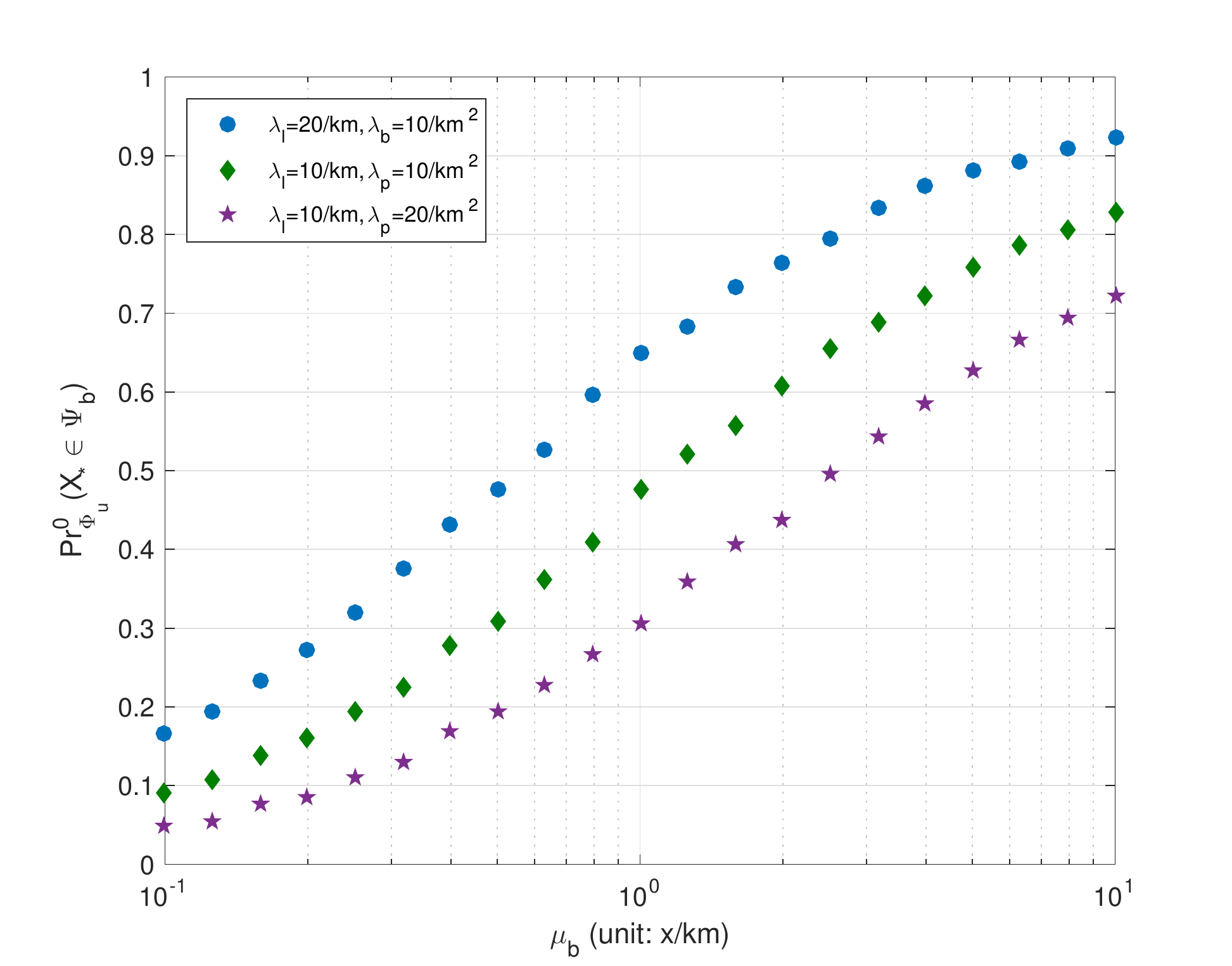}
	\caption{The probability that the typical planar user is associated with a vehicular base station on road. The horizontal axis is log-scaled.}
	\label{fig:associationofpoissonuser}
\end{figure}
Eq. \eqref{L4:1} or \eqref{L4:2} give the spatial average of the associations for Poisson distributed users.  
Fig. \ref{fig:associationofpoissonuser} illustrates the probability of the typical planar user being associated
with a vehicular base station. All three curves monotonically increase as $ \mu_b $ grows.
Notice the horizontal axis is log-scaled. The spatial intensity, $ \lambda_l\mu_b $,
varies from $ 1/\text{km}^2 $ to $ 100/\text{km}^2 $. Interestingly, when the spatial
intensities of the planar and vehicular base stations are exactly the same, i.e.,
in the middle curve, where $ \lambda_b=\lambda_l\mu_b=10/\text{km}^2$,
the typical planar user is slightly more likely to be associated with a planar base station. 

\begin{corollary}
Under the Palm distribution of $ \Phi_u, $ the distribution of the distance
from the origin to the nearest base station is given by
	\begin{align}
		f_{{R}}(r)=&2\pi\lambda_br e^{-\pi\lambda_br^2}\left(1-e^{-2\lambda_l\int_{0}^{r}1-e^{-2\mu_b\sqrt{r^2-t^2}\diff t}}\right)\nnb\\
		&+e^{-\pi\lambda_b r^2-2 \lambda_l \int_0^r 1-e^{-2 \mu_b \sqrt{r^2-u^2}} \, \diff u}\int_0^r \frac{4 \lambda_l \mu_b r e^{-2 \mu_b \sqrt{r^2-u^2}}}{\sqrt{r^2-u^2}}  \diff u.
	\end{align}
\end{corollary}
\begin{IEEEproof}
The proof immediately follows from the fact that the distance from the origin
to the nearest base station is given by the minimum of independent random variables $ R_p $ and $ R_v. $ 
\end{IEEEproof}
\subsection{Coverage Probability of the Typical Planar User}
The coverage probability under the Palm probability of $ \Phi_u $ is given by
$$
\bP_{\Phi_u}^0(\SINR>T)= 	\bP_{\Phi_u}^0(\SINR>T,X_\star\in\Phi_b)+\bP_{\Phi_u}^0(\SINR>T,X_\star\in\Psi_b).
$$
In the following, each term of the above expression is obtained in the form of an integral.
\begin{lemma}\label{Lemma:3}
The coverage probability by a planar base station is given by  
	\begin{align}
&\bP_{\Phi_u}^0(\SINR>T,X_\star\in\Phi_b)\nnb\\
&=\int_{0}^{\infty}2\pi\lambda_br e^{-\pi\lambda_br^2}e^{-2\pi\lambda_b\int_r^\infty\frac{Tr^\alpha u^{1-\alpha}}{1+Tr^\alpha u^{-\alpha}}\diff u}e^{-2\lambda_l\int_{r}^{\infty}1-e^{-2\mu_b\int_{0}^{\infty} \frac{Tr^\alpha (v^2+u^2)^{-\frac{\alpha}{2}}}{1+Tr^{\alpha}(v^2+u^2)^{-\frac{\alpha}{2}} } \diff u}\diff v}\nnb\\&\hspace{1.5cm}\exp\left({-2\lambda_l\int_{0}^{r}1-e^{-2\mu_b\sqrt{r^2-v^2}-2\mu_b\int_{\sqrt{r^2-v^2}}^{\infty} \frac{ Tr^\alpha (v^2+u^2)^{-\frac{\alpha}{2}}}{1+Tr^{\alpha}(v^2+u^2)^{-\frac{\alpha}{2}} } \diff u}\diff v}\right)\diff r\label{L5:1}.
	\end{align}
\end{lemma}
\begin{IEEEproof}
	The coverage probability by a planar base station is given by
	\begin{align}
		&\bP_{\Phi_u}^0(\SINR>T,X_\star\in\Phi_b)\nnb\\
		&=\bP_{\Phi_u}^0\left(\left.\frac{pH\|X_\star\|^{-\alpha}}{\sum_{X_k\in\Phi_b + \Psi_b \setminus \{X_\star\}} p H \|X_k\|^{-\alpha}}>T\right. , {X_\star\in\Phi_b}\right)\nnb\\
		&=\bE_{\Xi}\left[\bP_{\Phi_u}^0 \left(\left.H>T\|X_\star\|^{\alpha}{\sum_{X_k\in\Phi_b + \Psi_b  } H \|X_k\|^{-\alpha}},X_\star\in\Phi_b\right|\Xi\right)\right]\nnb\\
		&=\bE_{\Xi}\left[\int_{0}^{\infty} \bP_{\Phi_u}^0\left(\left.H>Tr^{\alpha}\!\!\!\!{\sum_{X_k\in\Phi_b + \Psi_b  }\!\!\!\!  H \|X_k\|^{-\alpha}},X_\star\in\Phi_b\right|\|X_\star\|=r,\Xi\right)f_{\|X_\star\|,X_\star\in\Phi_b|\Xi}(r)\diff r\right],\label{11}
		\end{align}
		where the integrand of Eq. \eqref{11} is 
		\begin{align}
			&\bP_{\Phi_u}^0\!\left(\left.H>Tr^{\alpha}\!\!\!\!{\sum_{X_k\in\Phi_b + \Psi_b  }\!\!\!\! H \|X_k\|^{-\alpha}},X_\star\in\Phi_b\right|\|X_\star\|=r,\Xi\right)\nnb\\
			&=\bP\left(\left.H>Tr^{\alpha}\left({\sum_{X_k\in\Phi_b} H \|X_k\|^{-\alpha}  \ind_{\|X_k\|>r}}+{\sum_{X_k\in\Psi_b} H \|X_k\|^{-\alpha}  \ind_{\|X_k\|>r}}\right)\right|\|X_\star\|,\Xi\right)\nnb.
		\end{align}
		Therefore, we have 
		\begin{align}
		&\bP_{\Phi_u}^0(\SINR>T,X_\star\in\Phi_b)\nnb\\&=\bE_{\Xi}\int_{0}^{\infty} \uta{\bE\left[\left.e^{-Tr^\alpha\!\!\!\sum\limits_{X_k\in\Phi_b}\!\!\! H \|X_k\|^{-\alpha}\ind_{\|X_k|>r}}\right.\right]}\utb{\bE\left[\left.e^{-Tr^\alpha\!\!\!\sum\limits_{X_k\in\Psi_b}\!\!\!H\|X_k\|^{-\alpha}\ind_{\|X_k|>r}}\right|\Xi\right]} f_{\|X_\star\|,X_\star\in\Phi_b|\Xi}(r)\diff r\nnb.
	\end{align}
\par For expression (a), the Laplace transform of the Poisson point process $ \Phi_b $ gives
\begin{equation}
\bE\left[\left.e^{-Tr^\alpha\sum\limits_{X_k\in\Phi_b} H \|X_k\|^{-\alpha} \ind_{\|X_k\|>r}}\right.\right]=e^{-2\pi\lambda_b\int_r^\infty\frac{Tr^\alpha u^{1-\alpha}}{1+Tr^\alpha u^{-\alpha}}\diff u}\label{16}.
\end{equation}
\par For expression (b), by using the explicit expressions for locations of the Cox points, $ X_{i,j} = (t_{j} \cos(\theta_i)-|r_i|\sin(\theta_i), t_{j}\sin(\theta_i)+|r_i|\cos(\theta_i)) $, we have 
\begin{align} 
&\bE\left[\left.e^{-Tr^\alpha\sum\limits_{X_k\in\Psi_b\setminus B(r)}H\|X_k\|^{-\alpha}\ind_{\|X_k\|>r}}\right|\Xi\right]\nnb\\
&=\bE\left[\left.\prod_{X_{i,j}\in\Psi_b}^{\|X_{i,j}\|>r } e^{-Tr^\alpha H_{i,j}  \|X_{i,j}\|^{-\alpha}}\right|\Xi\right]\nnb\\
&=\prod_{r_i\in\Xi}^{|r_i|\leq r}{\bE\left[\left.\prod_{t_j\in\phi_{l(r_i,\theta_i)}}^{t_{j}^2>r^2-r_i^2} e^{-Tr^\alpha H_{i,j} {(r_i^2+t_j^2)}^{-\frac{\alpha}{2}}}\right|\Xi\right]}\prod_{r_i\in\Xi}^{|r_i|> r}{\bE\left[\left.\prod_{t_j\in\phi_{l(r_i,\theta_i)}} e^{-Tr^\alpha H_{j}(r_i^2+t_j^2)^{-\frac{\alpha}{2}}}\right|\Xi\right]}\nnb\\
&=\prod_{r_i\in\Xi}^{|r_i|\leq r}\exp\left({-2\mu_b{\int_{{\sqrt{r^2-r_i^2}}}^{\infty}{\frac{Tr^\alpha(r_i^2+v^2)^{-\frac{\alpha}{2}}}{1+{Tr^\alpha (r_i^2+v^2)^{-\frac{\alpha}{2}}}}}\diff v}}\right)\nnb\\
&\hspace{0.5cm}\prod_{r_i\in\Xi}^{|r_i|> r}\exp\left({-2\mu_b\int_{0}^{\infty}\frac{Tr^\alpha(r_i^2+v^2)^{-\frac{\alpha}{2}}}{1+{Tr^\alpha (r_i^2+v^2)^{-\frac{\alpha}{2}}}}\diff v}\right)\label{17},
\end{align}
where we used the PGFL formula of the Poisson point processes on lines;
lines meeting the ball $ B(r) $ and lines avoiding this ball produce different expressions.
\par Finally, the distribution function of the distance from the origin to the nearest base station
with no other base stations exist inside  $ B(r) $ is given by 
\begin{align}
f_{\|X_\star\|}(r,X_\star\in\Phi_b|\Xi)&=\bP(\|X_\star\|=r,\Phi_b+\Psi_b(B(r))=0|\Xi)\nnb\\
&=\partial_r\left(1-\bP\left(\prod_{X_i\in\Phi_b}\ind_{\|X_\star\|\geq  r}\right)\right)\bP(\Psi_b(B(r))=0|\Xi)\nnnl
&=2\pi\lambda_br e^{-\pi\lambda_br^2}\prod_{r_i\in\Xi}^{|r_i|<r}\bE\left[\left.\prod_{t_j\in\phi_{l({r_i,\theta_i})}}\ind_{r_i^2+t_j^2>r^2}\right.\right]\nnb\\
&=2\pi\lambda_br e^{-\pi\lambda_br^2}\prod_{r_i\in\Xi}^{|r_i|<r}e^{-2\mu_b\sqrt{r^2-r_i^2}}\label{18'},
\end{align}
by independence of $ \Phi_b $ and $ \Psi_b. $
Combining  Eqs. \eqref{16}, \eqref{17}, and \eqref{18'}, we get 
\begin{align}
		&\bP_{\Phi_u}^0(\SINR>T,X_\star\in\Phi_b)\nnb\\
		&=\bE_{\Xi}\!\left[\!\int_{0}^{\infty}\!\!e^{-2\pi\lambda_b\int_r^\infty\frac{Tr^\alpha u^{1-\alpha}}{1+Tr^\alpha u^{-\alpha}}\diff u}\prod_{r_i\in\Xi}^{|r_i|\leq r}e^{-2\mu_b\int_{\sqrt{r^2-r_i^2}}^{\infty}\frac{Tr^\alpha(r_i^2+t^2)^{-\frac{\alpha}{2}}}{1+{Tr^\alpha (r_i^2+t^2)^{-\frac{\alpha}{2}}}}\diff t}\prod_{r_i\in\Xi}^{|r_i|> r}e^{-2\mu_b\int_{0}^{\infty}\frac{Tr^\alpha(r_i^2+t^2)^{-\frac{\alpha}{2}}}{1+{Tr^\alpha (r_i^2+t^2)^{-\frac{\alpha}{2}}}}\diff t}\right.\nnb\\
		&\hspace{2.0cm}\left.2\pi\lambda_br e^{-\pi\lambda_br^2}\prod_{r_i\in\Xi}^{|r_i|<r}e^{-2\mu_b\sqrt{r^2-r_i^2}}\diff r\right].\label{14}
\end{align}
Using Fubini's theorem, we have 
\begin{align}
&\bE_{\Xi}\left[\prod_{r_i\in\Xi}^{|r_i|\leq r}e^{-2\mu_b\sqrt{r^2-r_i^2}-2\mu_b\int_{\sqrt{r^2-r_i^2}}^{\infty}\frac{Tr^\alpha(r_i^2+t^2)^{-\frac{\alpha}{2}}}{1+{Tr^\alpha (r_i^2+t^2)^{-\frac{\alpha}{2}}}}\diff t}\prod_{r_i\in\Xi}^{|r_i|> r}e^{-2\mu_b\int_{0}^{\infty}\frac{Tr^\alpha(r_i^2+t^2)^{-\frac{\alpha}{2}}}{1+{Tr^\alpha (r_i^2+t^2)^{-\frac{\alpha}{2}}}}\diff t}\right]\nnb\\
&=e^{-2\lambda_l\int_{0}^{r}1-e^{-2\mu_b\sqrt{r^2-v^2}-2\mu_b\int_{\sqrt{r^2-v^2}}^{\infty} \frac{ Tr^\alpha (v^2+u^2)^{-\frac{\alpha}{2}}}{1+Tr^\alpha(v^2+u^2)^{-\frac{\alpha}{2}} } \diff u}\diff v}e^{-2\lambda_l\int_{r}^{\infty}1-e^{-2\mu_b\int_{0}^{\infty} \frac{Tr^\alpha (v^2+u^2)^{-\frac{\alpha}{2}}}{1+Tr^\alpha(v^2+u^2)^{-\frac{\alpha}{2}} } \diff u}\diff v}.\nnb
\end{align}
Incorporating it to Eq. \eqref{14} completes the proof.
\end{IEEEproof}
\begin{lemma}\label{Lemma:4}
The coverage probability by a vehicular  base station is given by  
	\begin{align}
	&\bP_{\Phi_u}^0(\SINR>T,X_\star\in\Psi_b)\nnb\\
	&=\bigintsss_0^{\infty}4\lambda_l\mu_bre^{-\pi\lambda_br^2}e^{-2\pi\lambda_b\int_r^\infty \frac{Tr^{\alpha}  u^{1-\alpha}}{1+Tr^{\alpha}  u^{-\alpha}}\diff u }\nnb\\
	&\hspace{1.5cm}\int_{0}^{\pi/2}e^{-2\mu_br\sin(\theta)-2\mu_b\int_{r\sin(\theta)}^{\infty}\frac{Tr^\alpha(r^2\cos^2(\theta)+v^2)^{-\frac{\alpha}{2}}}{1+Tr^\alpha(r^2\cos^2(\theta)+v^2)^{-\frac{\alpha}{2}}}\diff v}{\diff \theta}\nnb\\
	& \hspace{1.5cm}\exp\left(-2\lambda_l\int_{0}^{r} 1-e^{-2\mu_b\sqrt{r^2-u^2}}e^{-2\mu_b\int_{\sqrt{r^2-u^2}}^{\infty}\frac{Tr^\alpha(u^2+v^2)^{-\frac{\alpha}{2}}}{1+Tr^\alpha(u^2+v^2)^{-\frac{\alpha}{2}}}\diff v}\diff u\right)\nnb\\
	&\hspace{1.5cm}\exp\left(-2\lambda_l\int_{r}^{\infty}1-e^{-2\mu_b\int_{0}^{\infty}\frac{Tr^\alpha(u^2+v^2)^{-\frac{\alpha}{2}}}{1+{Tr^\alpha (u^2+v^2)^{-\frac{\alpha}{2}}}}\diff v}\diff u\right)\diff r.\label{L6:1}
	\end{align}
\end{lemma}
\begin{IEEEproof}
Notice that under the event $ X_\star\in\Psi_b $, the nearest base station is on a line.
Throughout the paper, we will denote this line and the point of $ \Xi $ giving this line
by $ l_\star $ and $ (r_\star,\theta_\star) $, respectively. 
To derive the coverage formula, let us first condition the coverage probability with respect
to the Poisson line process $ \Xi $, then  to the line $ l_\star $, and then to the
distance to the nearest base station $ \|X_\star\| $, sequentially. Then, we have 
	\begin{align}
	&\bP_{\Phi_u}^0(\SINR>T,X_\star\in\Psi_b)\nnb\\
	&=\bE_{\Xi}\bE_{l_\star}\left[\int_{0}^\infty \bP_{\Phi_u}^0\left(\left.H>Tr^{\alpha}\!\!\!\!\!\!{\sum_{X_k\in\Phi_b + \Psi_b +\phi_{l_\star(z)} }        \!\!\!\!\!\!  H \|X_k\|^{-\alpha}}, X_\star\in\Psi_b \right| \|X_\star\|=r,l_\star,\Xi\right) F(r)\diff r\right]\label{20''},
	\end{align}
where 
$F(r)=f_{\|X_\star\|,X_\star\in\Psi_b|l_\star,\Xi}(r) $,
the distribution of the distance from the origin to the nearest vehicular base station
on $ l_\star $ given $ l_\star $ and $ \Xi. $ In a similar way to the proof of Lemma 1,
the probability inside the triple integrals of \eqref{20''} is given by 
	\begin{align}
	&\bP_{\Phi_u}^0\left(\left.H>Tr^{\alpha}\!\!\!\!\!\!{\sum_{X_k\in\Phi_b + \Psi_b +\phi_{l_\star} }        \!\!\!\!\!\!  H \|X_k\|^{-\alpha}}, X_\star\in\Psi_b \right|\|X_\star\|=r,l_\star, \Xi\right)\nnb\\
	&=\bP\left(\left.H>Tr^{\alpha}\!\!\!\!\!\!{\sum_{X_k\in\Phi_b + \Psi_b +\phi_{l_\star} }        \!\!\!\!\!\!  H \|X_k\|^{-\alpha}}\ind_{\|X_k\|>r}\right|\|X_\star\|=r,l_\star,\Xi\right).\nnb
	\end{align}
Therefore, it is described by the product of the following expressions 
	\begin{align}
	\bE \left[\left.e^{-Tr^\alpha\!\!\!\sum\limits_{X_k\in\Phi_b}\!\!\! H_k \|X_k\|^{-\alpha} \ind_{\|X_k\|>r} }\right.\right]=&e^{-2\pi\lambda_b\int_r^\infty \frac{Tr^\alpha  u^{1-\alpha}}{1+Tr^\alpha  u^{-\alpha}}\diff u}.\label{20'}\\
	\bE\left[\left.e^{-Tr^\alpha\!\!\!\sum\limits_{X_k\in\Psi_b}\!\!\!H_k\|X_k\|^{-\alpha} \ind_{\|X_k\|>r}}\right|\Xi\right]=&\prod_{r_i\in\Xi}^{|r_i|\leq r}e^{-2\mu_b\int_{\sqrt{r^2-r_i^2}}^{\infty}\frac{Tr^\alpha(r_i^2+v^2)^{-\frac{\alpha}{2}}}{1+{Tr^\alpha (r_i^2+v^2)^{-\frac{\alpha}{2}}}}\diff v}\nnb\\
	&\prod_{r_i\in\Xi}^{|r_i|> r}e^{-2\mu_b\int_{0}^{\infty}\frac{Tr^\alpha(r_i^2+v^2)^{-\frac{\alpha}{2}}}{1+{Tr^\alpha (r_i^2+v^2)^{-\frac{\alpha}{2}}}}\diff v}.\label{23}\\
	\bE\left[\left.\prod_{t_j\in \phi_{l_\star(z)}}^{z^2+t_j^2>r^2} e^{-Tr^\alpha {(z^2+t_j^2)}^{-\frac{\alpha}{2}}}\right|(r_\star,\theta_\star)=(z,\theta)\right]
	=&\left.e^{-2\mu_b \int_{\sqrt{r^2-z^2}}^\infty\frac{Tr^\alpha(z^2+v^2)^{-\frac{\alpha}{2}}}{1+Tr^\alpha(z^2+v^2)^{-\frac{\alpha}{2}}}\diff v}\right.\label{20},
	\end{align}
where Eq. \eqref{20} is obtained by the Laplace transform of the Poisson point
process on the line of parameter $ (z,\theta) $ with $ |z|<r. $
	\par  The distribution function $F(r) $ of Eq. \eqref{20''} is the probability
that line $ l_\star $ has a point at a distance $ r $ and the point process
$ \Psi_{b}+\Phi_b+\phi_{l_\star(z)} $ is empty of points inside $ B(r) $. In other words,
	\begin{align}
	&f_{\|X_\star\|,X_\star\in\Psi_b|l_\star,\Xi}(r)\nnb
	\\
	&=\bP(\|X_\star\|=r,\Phi_b+\Psi_b(B(r))=0|l_\star,\Xi)\nnb\\
	&=\bP\left(\|X_\star\|=r,\Phi_b(B(r))=0,\Psi^{!l_\star}(B(r))=0\right)\nnb\\
	&=\partial_r(1-\bP(\phi_{l_\star(z)}(B(r))=0|l_\star))\bP(\Phi_b(B(r))=0)\bP(\Psi_b(B(r))=0|\Xi)\nnb\\
	&=\partial_r(1-e^{-2\mu_b\sqrt{r^2-z^2}})e^{-\pi\lambda_b r^2}\prod_{r_i\in\Xi}^{|r_i|<r}e^{-2\mu_b\sqrt{r^2-r_i^2}}\nnb\\
	&=\frac{2\mu_b re^{-2\mu_b\sqrt{r^2-z^2}}}{\sqrt{r^2-z^2}}e^{-\pi\lambda_b r^2}\prod_{r_i\in\Xi}^{|r_i|<r}e^{-2\mu_b\sqrt{r^2-r_i^2}}\label{26},
	\end{align}
where we used independence and Slivnyak's theorem.
	\par Finally, we apply Fubini's theorem to evaluate the triple integral in \eqref{20''}.
By changing the order of integrals, the expectation with respect to $ l_\star $ is evaluated first.
By Campbell's averaging formula, we have 
	\begin{align}
	&\bE_{l_\star(z)}\left[\frac{2\mu_br e^{-2\mu_b\sqrt{r^2-z^2}}}{\sqrt{r^2-z^2}}\exp\left({-2\mu_b \int_{\sqrt{r^2-z^2}}^\infty\frac{Tr^\alpha(z^2+v^2)^{-\frac{\alpha}{2}}}{1+Tr^\alpha(z^2+v^2)^{-\frac{\alpha}{2}}}\diff v}\right)\right]\nnb\\
	&=\int_{0}^r\frac{4\lambda_l\mu_br  e^{-2\mu_b\sqrt{r^2-z^2}}}{\sqrt{r^2-z^2}}\exp\left({-2\mu_b \int_{\sqrt{r^2-z^2}}^\infty\frac{Tr^\alpha(z^2+v^2)^{-\frac{\alpha}{2}}}{1+Tr^\alpha(z^2+v^2)^{-\frac{\alpha}{2}}}\diff v}\right) \diff z\nnb\\
	&=\int_{0}^{\pi/2}4\lambda_l\mu_be^{-2\mu_br\sin(\theta)-2\mu_b\int_{r\sin(\theta)}^{\infty}\frac{Tr^\alpha(r^2\cos^2(\theta)+v^2)^{-\frac{\alpha}{2}}}{1+Tr^\alpha(r^2\cos^2(\theta)+v^2)^{-\frac{\alpha}{2}}}\diff v}{\diff \theta}.\label{27''}
	\end{align}
Then, the integral with respect to the Poisson point process $ \Xi $  gives
	\begin{align}
	&\bE_{\Xi}\left[\prod_{r_i\in\Xi}^{|r_i|\leq r}{ e^{-2\mu_b\sqrt{r^2-r_i^2}}e^{-2\mu_b\int_{\sqrt{r^2-r_i^2}}^{\infty}\frac{Tr^\alpha(r_i^2+v^2)^{-\frac{\alpha}{2}}}{1+{Tr^\alpha (r_i^2+v^2)^{-\frac{\alpha}{2}}}}\diff v} }{}\prod_{r_i\in\Xi}^{|r_i|> r}e^{-2\mu_b\int_{0}^{\infty}\frac{Tr^\alpha(r_i^2+v^2)^{-\frac{\alpha}{2}}}{1+{Tr^\alpha (r_i^2+v^2)^{-\frac{\alpha}{2}}}}\diff v}\right]\nnb
	\\	&=\exp\left(-2\lambda_l\int_{0}^{r} 1-e^{-2\mu_b\sqrt{r^2-u^2}}e^{-2\mu_b\int_{\sqrt{r^2-u^2}}^{\infty}\frac{Tr^\alpha(u^2+v^2)^{-\frac{\alpha}{2}}}{1+Tr^\alpha(u^2+v^2)^{-\frac{\alpha}{2}}}\diff v}\diff u\right)\nnb\\
	&\hspace{0.5cm}\exp\left(-2\lambda_l\int_{r}^{\infty}1-e^{-2\mu_b\int_{0}^{\infty}\frac{Tr^\alpha(u^2+v^2)^{-\frac{\alpha}{2}}}{1+{Tr^\alpha (u^2+v^2)^{-\frac{\alpha}{2}}}}\diff v}\diff u\right)\label{28},
	\end{align}
where we use Slivnyak's theorem and the PGFL formula.
Combining Eqs. \eqref{20'}, \eqref{27''} and \eqref{28} completes the proof.
Another version of proof is provided in Appendix \ref{A:1}	
\end{IEEEproof}
Fig. \ref{fig:coveragepoissonuser} illustrates the coverage probability of the typical planar user.
We first consider parameters $ \lambda_l=5.34/\text{km},\lambda_b=6.15/\text{km}^2 $, and $ \mu_b=5/\text{km} $. They are proposed to illustrate the 3GPP LTE urban V2X scenario where inter site distance of hexagonal base station is 500 meters and road block of Manhattan grid is 250 meters by 433 meters\footnote{In the proposed model, the mean area of surface surrounded by Poisson lines is $ \pi/\lambda_l^2$. On the other hand, assuming no vehicular base stations, the mean area of Poisson-Voronoi cell of macro base station is $ 1/\lambda_b$\cite{chiu2013stochastic}. Therefore, inter site distance of 500 meters is translated into $ \lambda_b=6.15 $ while the road block of size 250 meters by 433 meters is translated into $ \lambda_l=5.3. $}\cite{3gpp36885}. 
In this setting, the coverage probability by vehicular base stations significantly higher than the one provided by planar base stations. It elaborates the use of vehicle-to-planar user (i.e., planar pedestrians) communications since those users have higher coverage by vehicular base stations than by planar base stations. 
\par We also evaluate the coverage probability when the densities of vehicular base stations and planar base stations are the same: $ \lambda_l=5/\text{km},  \mu_b=5/\text{km},$ and $ \lambda_b=25/\text{km}^2 $. In this case, the impact of topological differences on the coverage probability is emphasized. Unexpectedly, planar users still have a slightly higher coverage probability by vehicular base stations. In the specific scenario, the average distance from the typical planar user to its nearest vehicular base station is 85 meters while it increases up to 125 meters with its nearest planar base station.
\begin{figure}
	\centering
	\includegraphics[width=0.5\linewidth]{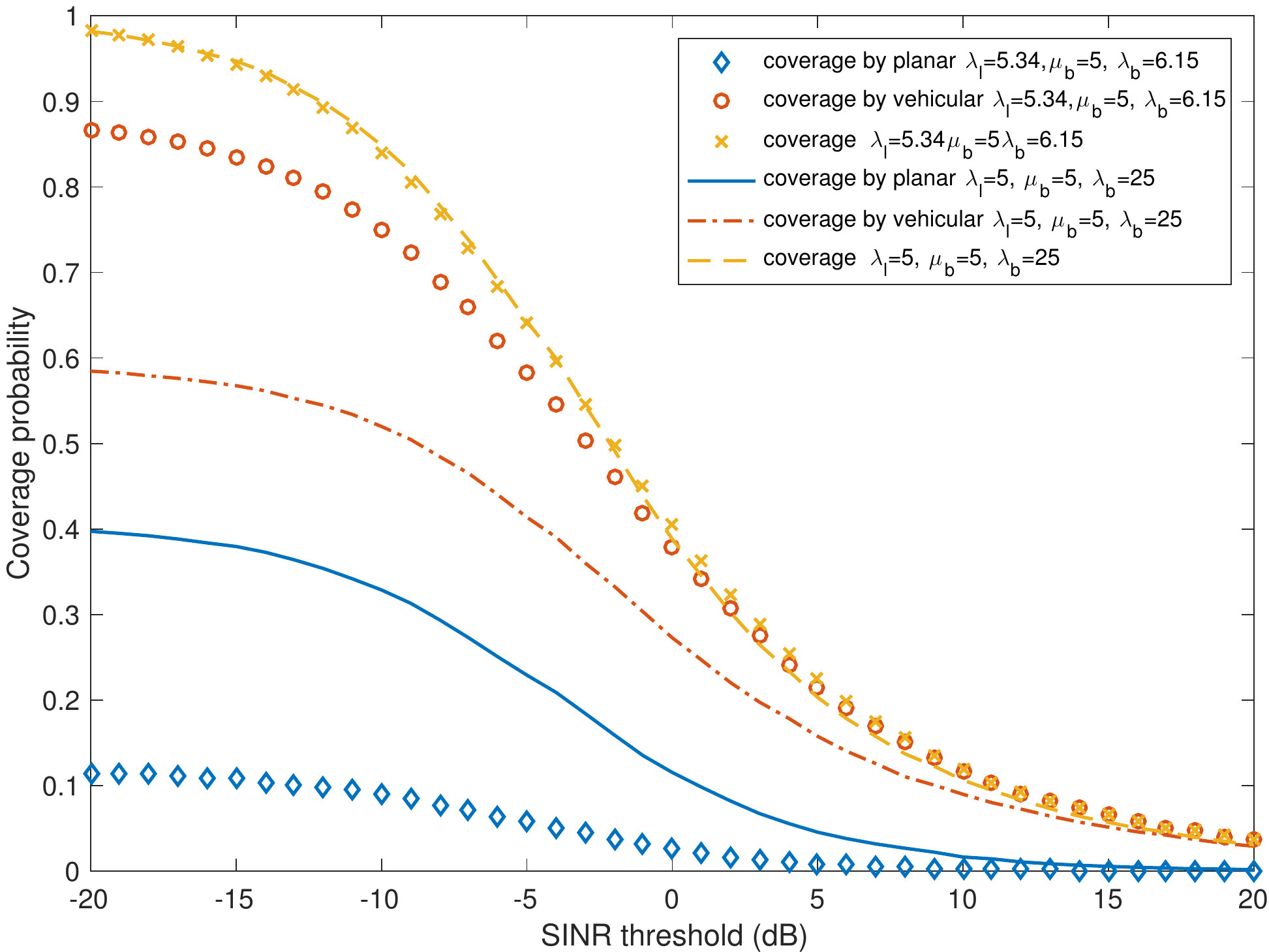}
	\caption{The coverage probability of the typical planar user under the Palm distribution of $ \Phi_u $.}
	\label{fig:coveragepoissonuser}
\end{figure}

\section{Coverage: Palm Probability of $ {\Psi_u} $}\label{S:4}
In this section, we derive the user association and coverage probabilities under
the Palm distribution $ \bP_{\Psi_u}^0. $ Note that under this probability,
the typical vehicular user is located at the origin and there is a road passing
through the origin and containing the typical user.
As in Section \ref{S:3}, Slivnyak's theorem is very useful to produce integral formulas.
\subsection{Association of the Typical Vehicular User}
The typical user is associated with either a planar base station of $ \Phi_b $
or a vehicular base station of $ \Psi_b. $ These events are denoted by 
$ X_\star\in\Phi_b $ and $ X_\star\in\Psi_b, $ respectively. 
\begin{proposition}\label{Prop:2}
Under the Palm distribution $ \bP_{\Psi_u}^0, $ the probabilities of the typical
vehicular user to be associated with a planar base station and a vehicular base station are given by 
	\begin{align}
	\bP_{\Psi_u}^0(X_\star\in\Phi_b)&=\int_{0}^{\infty}2\pi\lambda_b w e^{-\pi\lambda_b w^2-2\mu_b w -2\lambda_l\int_0^w 1-e^{-2\mu_b\sqrt{w^2-t^2}}\diff t}\diff w,\label{L7:1}\\
	\bP_{\Psi_u}^0(X_\star\in\Psi_b)&=1-\int_{0}^{\infty}2\pi\lambda_b w e^{-\pi\lambda_b w^2-2\mu_b w - 2\lambda_l\int_0^w 1-e^{-2\mu_b\sqrt{w^2-t^2}}\diff t}\diff w,\label{L7:2}
	\end{align}
	respectively.
\end{proposition}
\begin{IEEEproof}
Let us denote the line containing the origin and the Poisson process on the
line by $ l_{0} $ and $ \phi_{l_0} $, respectively. Under the Palm distribution of $ {\Psi_u}, $ let us define 
$$ \hat{R_p}=\inf_{X_k\in\Phi_b}\|X_k\| \text{ and } \hat{\hat{R_v}}=\inf_{X_k\in\Psi_b+\phi_{l_0}}\|X_k\| $$
in order to describe the distances to the nearest points of $ \Phi_b $ and $ \Psi_b $, respectively.
As in Proposition 1, the association probability is given by  
	\begin{align}
	\bP_{\Psi_u}^0(X_\star\in\Phi_b)&=\bE_{\Psi_u}^0\left[\ind_{\hat{R_p}<\hat{R_v}}\right]=\int_{0}^{\infty}\bP_{\Psi_u}^0(\hat{R_v}>r) f_{\hat{R_p}}(r)\diff r,\label{22}
	\end{align}
where the integrand is given as follows:
	\begin{align}
	\bP_{\Psi_u}^0(\hat{R_v}>r_p)&=\bE\left[\prod_{X_k\in\Psi_b+\phi_{l_0}}\ind_{\|X_k\|>r}\right]\nnb\\
	&=\bE\left[\prod_{t_j\in\phi_{l_0}}\ind_{|t_{j}|>r_p}\right]\bE\left[\prod_{r_i\in\Xi}^{|r_i|<r}\bE\left[\prod_{t_j\in\phi_{l(r_i,\theta_i)}}\ind_{t_{j}^2>r_p^2-r_i^2}|\Xi\right]\right]\nnb\\
	&=e^{-2\mu_b r_p} e^{-2\lambda_l\int_{0}^{r_p}1-e^{-2\mu_b \sqrt{r_p^2-t^2}} \diff t}\label{23''}.
	\end{align}
Due to the independence property of Poisson point processes, the distribution function of $ \hat{R_p} $ is equal to 
	\begin{equation}
	f_{\hat{R_p}}(w)= 2 \pi\lambda_b  w e^{-\pi \lambda_bw^2 }\label{24}.
	\end{equation}
Combining Eqs. \eqref{22}, \eqref{23''} and \eqref{24} gives the result. The proof is concluded by observing that
$ \bP_{\Psi_u}^{0}(X_\star\in\Psi_b)=1-\bP_{\Psi_u}^{0}(X_\star\in\Phi_b)$.
\end{IEEEproof}
\begin{corollary}
Under the Palm distribution $ \bP_{\Psi_u}^0, $ the distribution of the distance
from the origin to the nearest base station is given by 
	\begin{align}
	f_{{R}}(r)=&2\pi\lambda_br e^{-\pi\lambda_br^2}e^{-2\mu_br}\left(1-e^{-2\lambda_l\int_{0}^{r}1-e^{-2\mu_b\sqrt{r^2-u^2}\diff u}}\right)\nnb\\
	&+e^{-\pi\lambda_b r^2}2\mu_be^{-2\mu_br}\left(1-e^{-2 \lambda_l \int_0^r 1-e^{-2 \mu_b \sqrt{r^2-u^2}} \, \diff u}\right)\nnb\\
	&+e^{-\pi\lambda_b r^2}e^{-2\mu_br}e^{-2 \lambda_l \int_0^r 1-e^{-2 \mu_b \sqrt{r^2-u^2}} \, \diff u}\int_0^r \frac{4 \lambda_l \mu_b r e^{-2 \mu_b \sqrt{r^2-u^2}}}{\sqrt{r^2-u^2}}  \diff u.
	\end{align}
\end{corollary}
\begin{IEEEproof}
The proof immediately follows from the fact that, under the Palm distribution of $ \Psi_u,$
the distance from the origin to the nearest point of $ \Phi_b+\Psi_b $ is the minimum of
the two {independent} random variables $ \hat{R_p}$ and $ \hat{R_v} $.
\end{IEEEproof}
Note that the distribution functions of the nearest distance are different under $ \bP_{\Phi_u}^0 $ and $ \bP_{\Psi_u}^0 $. 

\subsection{Coverage Probability of the Typical Vehicular User}
Since $ (X_\star\in\Psi_b)=(X_\star\in\Phi_b)^c $, the coverage probability is described as follows.
\begin{align}
\bP_{\Psi_u}^0(\SINR>T)=& \bP_{\Psi_u}^0(\SINR>T,X_\star\in\Phi_b)+\bP_{\Psi_u}^0(\SINR>T,X_\star\in\Psi_b).\nnb
\end{align}
In the following, each term is derived as an integral formula. 

\begin{lemma}\label{Lemma:5}
Under the Palm distribution of $ {\Psi_u}, $ the coverage probability by a planar base station is given by  
\begin{align}
&\bP_{\Psi_u}^0(\SINR>T,X_\star\in\Phi_b)\nnb\\
&=\int_{0}^{\infty}\!\!e^{-2\mu_b\int_{r}^{\infty} \frac{Tr^{\alpha} u^{-{\alpha}}}{1+Tr^{\alpha}u^{-{\alpha}}}\diff u}e^{-2\pi\lambda_b\int_{r}^{\infty}\frac{Tr^{\alpha} u^{1-\alpha}}{1+Tr^{\alpha}u^{-\alpha}}\diff u}e^{-2\lambda_l\int_{0}^{r}1-e^{-2\mu_b\int_{\sqrt{r^2-v^2}}^{\infty} \frac{Tr^{\alpha} (v^2+u^2)^{-\frac{\alpha}{2}}}{1+Tr^{\alpha}(v^2+u^2)^{-\frac{\alpha}{2}} } \diff u}\diff v } \nnnl
&\hspace{1.2cm}2\pi\lambda_br e^{-\pi\lambda_b r^2-2\mu_b r}\exp\left({-2\lambda_l\int_{0}^{r}1-e^{-2\mu_b\sqrt{r^2-v^2}-2\mu_b\int_{\sqrt{r^2-v^2}}^{\infty} \frac{ Tr^\alpha (v^2+u^2)^{-\frac{\alpha}{2}}}{1+Tr^\alpha(v^2+u^2)^{-\frac{\alpha}{2}} } \diff u}\diff v}\right)\diff r. \label{L8:1}
\end{align}
\end{lemma}
\begin{IEEEproof}
Conditionally on the Poisson point process $ \Xi $ and on the distance to the 
nearest base station $ \|X_\star\| $, the coverage probability under the Palm distribution of $ \Psi_u $ is given by 
		\begin{align}
	&\bP_{\Psi_u}^0(\SINR>T,X_\star\in\Phi_b)\nnb\\			&=\bE\int_{0}^{\infty} \bP_{\Psi_u}^0\left(\left.H>Tr^{\alpha}\!\!\!\!\!{\sum_{X_k\in\Phi_b + \Psi_b}\!\!\!\!\!  H \|X_k\|^{-\alpha}},X_\star\in\Psi_b \right|\|X_\star\|=r,\Xi\right)f_{\|X_\star\,X_\star\in\Psi_b|\Xi}(r)\diff r\nnb\\
			&=\bE\int_{0}^{\infty} \bP\left(\left.H>Tr^{\alpha}\!\!\!\!\!\!\!\!\!{\sum_{X_k\in\Phi_b + \Psi_b+\phi_{l_0} }\!\!\!\!\!\!\!\!\!  H \|X_k\|^{-\alpha}},X_\star\in\Psi_b \right|\|X_\star\|=r,\Xi\right)f_{\|X_\star\,X_\star\in\Psi_b|\Xi}(r)\diff r\nnb\\
&=\bE\int_{0}^{\infty} \uta{\bE\left[\left.e^{-Tr^\alpha\!\!\!\sum\limits_{X_k\in\Phi_b}\!\!\! H \|X_k\|^{-\alpha}\ind_{\|X_k|>r}}\right.\right]}\utb{\bE\left[\left.e^{-Tr^\alpha\!\!\!\sum\limits_{X_k\in\Psi_b}\!\!\!H\|X_k\|^{-\alpha}\ind_{\|X_k|>r}}\right|\Xi\right]} \nnb\\
&\hspace{1.74cm}{\bE\left[\left.e^{-Tr^\alpha\!\!\!\!\sum\limits_{X_k\in\phi_{l_0}}\!\!\!\!H\|X_k\|^{-\alpha}\ind_{\|X_k|>r}}\right.\right]}f_{\|X_\star\|,X_\star\in\Psi_b|\Xi}(r)\diff r\label{23'},
	\end{align}
where (a) and (b) are given by Eq. \eqref{20'} and \eqref{23}, respectively. We have  
\begin{align}
		\bE\left[\left.e^{-Tr^\alpha\!\!\!\sum\limits_{X_k\in\phi_{l_0}}\!\!\!H\|X_k\|^{-\alpha}\ind_{\|X_k|>r}}\right.\right]=&e^{-2\mu_br\int_{r}^{\infty}\frac{Tr^\alpha v^{-\alpha}}{1+Tr^\alpha v^{-\alpha}}\diff v},\label{29}
	\end{align}
using the Laplace transform of the Poisson point process $ \phi_{l_0}, $
which is independent of $ \Xi. $
\par Under the Palm distribution of $ \Psi_u $, the distribution of 
the distance from the origin to the nearest base station is given by  
\begin{align}
f_{\|X_\star\|,X_\star\in\Phi_b|\Xi}(r)&=\bP_{\Psi_u}^0\left(\|X_\star\|=r,X_\star\in\Phi_b|\Xi\right)\nnb\\
&=\partial_r(1-\bP_{\Psi_u}^0(\Phi_b(B(r))=0))\bP_{\Psi_u}^0(\Psi_b(B(r))=0|\Xi)\nnb\\
&=2\pi\lambda_br e^{-\pi\lambda_b r^2}\bP(\Psi_b^{l_0}(B(r))=0)\nnb\\
&=2\pi\lambda_br e^{-\pi\lambda_b r^2}e^{-2\mu_b}\prod_{r_i\in\Xi}^{|r_i|<r}e^{-2\mu_b\sqrt{r^2-r_i^2}},\label{39}
\end{align}
where the last expression is given by Slivnyak's theorem and the independence property:
\begin{align}
	\bP({\Psi_b}^{l_0}(B(r))=0|\Xi)&=\bP(\Psi_b^{!{l_0}}(B(r))=0|\Xi)\bP(\phi_{l_0}(B(r))=0|\Xi)\nnb\\
	&=e^{-2\mu_br}\prod_{r_i\in\Xi}^{|r_i|<r}e^{-2\mu_b\sqrt{r^2-r_i^2}}.\nnb
\end{align}
Combining Eqs. \eqref{20'}, \eqref{23}, \eqref{29}, and \eqref{39} and applying Fubini's
theorem allows one to complete the proof.
\end{IEEEproof}
\begin{lemma}\label{Lemma:6}
Under the Palm probability of $ \Psi_u, $ the coverage probability by a vehicular base station is given by  
	\begin{align}
		\bP_{\Psi_u}^0(\SINR>T,X_\star\in\Psi_b)=&\uta{\bP_{\Psi_u}^0(\SINR>T,X_\star\in\Psi_b,l_0\equiv l_\star )}\nnnl&+\utb{\bP_{\Psi_u}^0(\SINR>T,X_\star\in\Psi_b,l_0\not\equiv l_\star)},\label{L9:1}
	\end{align}
where  (a) and (b) are given by Eqs. \eqref{31''} and \eqref{part2}, respectively.
More precisely, part (a) gives the probability that the typical vehicular user is covered and its
nearest vehicular base station is on its road, whereas part (b) gives the 
probability that the typical vehicular user is covered and its nearest
vehicular base station is not on its road.
\end{lemma}
\begin{IEEEproof} Under the Palm probability and under the event $ \{X_\star\in\Psi_b\} $,
we can consider two lines that are not necessarily distinct: the one containing the origin, $ l_0 $,
and the one containing the nearest base station $ l_\star $.
The coverage probability is then partitioned into the events $ \{l_\star\equiv l_0 \} $
and  $ \{l_\star \not\equiv l_0\}, $ which are denoted by $ \cE_1 $ and $ \cE_2 $, respectively. Then, we have 
	\begin{align}
\bP_{\Psi_u}^0(\SINR>T,\cV)={\bP_{\Psi_u}^0(\SINR>T,\cV,\cE_1)}
&+{\bP_{\Psi_u}^0(\SINR>T,\cV,\cE_2)},\nnb
	\end{align}
where $ \cV $ denotes the event $ X_\star\in\Psi_b $.
\par In order to get an expression for (a), we condition on the Poisson line process.
Then, the coverage probability is given by 
\begin{align}
&{\bP_{\Psi_u}^0(\SINR>T ,\cV,\cE_1 )}\nnb\\
&=\bE_{\Xi}\!\bigintsss_{0}^{\infty} {\bE\left[\left.e^{-Tr^\alpha \!\!\!\sum\limits_{X_{k}\in\Phi_b}\!\!\!H_{k}\|X_{k}\|^{-{\alpha}}\ind_{\|X_i\|>r}}\right.\right]}{\bE\left[\left.e^{-Tr^\alpha\!\!\!\sum\limits_{X_{k}\in\Psi_b}\!\!\! H_k\|X_{k}\|^{-\alpha} \ind_{\|X_k\|>r}}\right|\Xi \right]}\nnb\\
&\hspace{2.0cm}\bE\left[\left.e^{-Tr^\alpha \sum\limits_{X_{k}\in\phi_{l_0}}\!\!\!H_{k}\|X_{k}\|^{-{\alpha}}\ind_{\|X_i\|>r}}\right.\right]f_{\|X_\star\|,\cV,\cE_1|\Xi}(r)\diff r ,\label{31'''} 
\end{align}
where the integrand is already given as the product of Eqs. \eqref{20'}, \eqref{23} and \eqref{29}.
Analogous to the distribution function given in the proof of Lemma 3, the above distribution function is 
	\begin{align}
f_{\|X_\star\|,\cV,\cE_1|\Xi}&=\bP_{\Psi_u}^0(\|X_\star\|=r,X_\star\in\Psi_b,\Phi_b+\Psi_b(B(r))=0|\Xi)\nnb\\
&=\bP_{\Psi_u}^{l_0}(\|X_\star\|=r,X_\star\in\phi_{l_0},\Phi_b+\Psi_b^{!l_0}(B(r))=0|\Xi)\nnb\\
&=\partial_r\left(\bP\left(1-\prod_{X_i\in\phi_{l_0}}\ind_{\|X_i\|\geq r}\right)\right)\bP_{\Psi_u}^{l_0}(\Phi_b+\Psi_b^{!l_0}(B(r))=0|\Xi)\nnb\\
&=\partial_r\left(1-\exp(-2\mu_b r)\right)\cdot\bP(\Phi_b(B(r))=0|\Xi)\cdot\bP(\Psi_{b}(B(r))=0|\Xi)\nnb\\
&=2\mu_b e^{-2\mu_br}e^{-\pi\lambda_b r^2}\prod_{r_i\in\Xi}^{|r_i|<r}e^{-2\mu_b\sqrt{r^2-r_i^2}}\label{32''}.
\end{align}
Incorporating \eqref{20'} \eqref{23}, \eqref{29}, and \eqref{32''} into Eq. \eqref{31'''} and applying Fubini's theorem gives
\begin{align}
	&{\bP_{\Psi_u}^0(\SINR>T ,X_\star\in\Phi_b,l_0\equiv l_\star )}\nnb\\
&=\bigintsss_{0}^{\infty}\!\! e^{-2\pi\lambda_b\int_{r}^{\infty}\frac{Tr^{\alpha} u^{1-\alpha}}{1+Tr^{\alpha}u^{-\alpha}}\diff u-2\mu_b\int_{r}^{\infty} \frac{Tr^{\alpha} u^{-{\alpha}}}{1+Tr^{\alpha}u^{-{\alpha}}}\diff u-2\lambda_l\int_{r}^{\infty}1-e^{-2\mu_b\int_{0}^{\infty} \frac{Tr^{\alpha} (v^2+u^2)^{-\frac{\alpha}{2}}}{1+Tr^{\alpha}(v^2+u^2)^{-\frac{\alpha}{2}} } \diff u}\diff v}\nnb\\
	& \hspace{1.3cm}2\mu_b e^{-2\mu_b r-\pi\lambda_br^2}\exp\left({-2\lambda_l\int_{0}^{r}1-e^{-2\mu_b\sqrt{r^2-v^2}-2\mu_b\int_{\sqrt{r^2-v^2}}^{\infty} \frac{ Tr^\alpha (v^2+u^2)^{-\frac{\alpha}{2}}}{1+Tr^\alpha(v^2+u^2)^{-\frac{\alpha}{2}} } \diff u}\diff v}\right) \diff r.\label{31''}
\end{align}
	\par On the other hand, for expression (b) of Eq. \eqref{L9:1},
conditionally on $ \Xi^{} $ and on $ l_\star $, the coverage probability is given by   
\begin{align}
&\bP_{\Psi_u}^0(\SINR>T,\cV,\cE_2 )\nnb\\
&=\bE_{\Xi}\bE_{l_\star}\int_{0}^{\infty}\bP\left(\left.H>Tr^\alpha\!\!\!\!\!\!\!\!\!\!\!\!\!\!\!\!\sum_{X_k\in\Phi_b + \Psi_b +\phi_{l_\star(z)} +\phi_{l_0}}\!\!\!\!\!\!\!\!\!\!\!\!\!\!\!\!H_k\|X_k\|^{-\alpha}\ind_{\|X_k\|>r},\cV,\cE_2\right|\|X_\star\|,l_\star,\Xi\right)F(r)\diff r,
\end{align}
where the integrands are given by 
\begin{align}
&\bP\left(\left.H>Tr^\alpha\!\!\!\!\!\!\!\!\!\!\!\!\sum_{X_k\in\Phi_b + \Psi_b +\phi_{l_\star(z)} +\phi_{l_0}}\!\!\!\!\!\!\!\!\!\!\!\!H_k\|X_k\|^{-\alpha}\ind_{\|X_k\|>r},\cV,\cE_2\right|\|X_\star\|,l_\star,\Xi\right)\nnb\\
&=\bE\left[\left.e^{-Tr^\alpha \sum\limits_{X_{k}\in\Phi_b} H_k\|X_{k}\|^{-\alpha}\ind_{\|X_k\|>r}}\right. \right]\bE\left[\left.e^{-Tr^\alpha \sum\limits_{X_{k}\in\Psi_b} H_k\|X_{k}\|^{-\alpha}\ind_{\|X_k\|>r}}\right| \Xi\right]\nnb\\
&\hspace{0.5cm}\bE\left[\left.e^{-Tr^\alpha \sum\limits_{X_{k}\in\phi_{l_0}}\!\!\! H_k\|X_{k}\|^{-\alpha}\ind_{\|X_k\|>r}}\right.\right]\bE\left[\left.e^{-Tr^\alpha \sum\limits_{X_{k}\in\phi_{l_\star}}\!\! H_k\|X_{k}\|^{-\alpha}\ind_{\|X_k\|>r}}\right|l_\star\right]\nnb,
\end{align}
that are already given by Eqs. \eqref{20'}, \eqref{23}, \eqref{29}, and \eqref{20}, respectively. The function $ F(r) $ is  
\begin{align}
	F(r)&=\bP_{\Psi_u}^0(\|X_\star\|=r,X_\star\in\phi_{l_\star},\Phi_b+\Psi_b(B(r))=0|\Xi,l_\star)\nnb\\
	&=\bP_{\Psi_u}^0(\|X_\star\|=r,X_\star\in\phi_{l_\star}|l_\star(z,\theta))\bP_{\Psi_u}^0\left(\Phi_b+\Psi_b(B(r))=0|\Xi\right)\nnb\\
	&=\partial_r\left(1-\bP\left(\left. \prod_{X_{i\in\phi_\star}}\ind_{\|X_i\|\geq r}\right|l_\star(z,\theta)\right)\right)\bP_{\Phi_u}^{l_0}(\phi_{l_0}+\Phi_b+\Psi_b^{!l_0}(B(r))=0|\Xi)\nnb\\
	&=\frac{2\mu_b re^{-2\mu_b\sqrt{r^2-z^2}}}{\sqrt{r^2-z^2}}\bP(\phi_l(B(r))=0)\bP(\Phi_b(B(r))=0)\bP(\Psi_b(B(r))=0|\Xi)\nnb\\
	&=\frac{2\mu_b re^{-2\mu_b\sqrt{r^2-z^2}}}{\sqrt{r^2-z^2}}e^{-2\mu_b r}e^{-\pi\lambda_b r^2}\prod_{r_i\in\Xi}^{|r_i|<r}e^{-2\mu_b\sqrt{r^2-r_i^2}}\nnb.
\end{align}
Therefore, we combine the above integrands and the distribution function and apply Fubini's theorem to have 
\begin{align}
	&\bP_{\Phi_u}^0(\SINR>T,X_\star\in\Psi_b,l_0\not\equiv l_\star)\nnb\\
&=\bigintsss_0^{\infty}4\lambda_l \mu_bre^{-2\mu_br-\pi\lambda_br^2-2\pi\lambda_b\int_r^\infty \frac{Tr^{\alpha}  u^{1-\alpha}}{1+Tr^{\alpha}  u^{-\alpha}}\diff u -2\mu_b\int_r^\infty \frac{Tr^{\alpha}  u^{-\alpha}}{1+Tr^{\alpha}  u^{-\alpha}}\diff u }\nnb\\
&\hspace{1.5cm}\int_{0}^{\pi/2}e^{-2\mu_br\sin(\theta)-2\mu_b\int_{r\sin(\theta)}^{\infty}\frac{Tr^\alpha(r^2\cos^2(\theta)+v^2)^{-\frac{\alpha}{2}}}{1+Tr^\alpha(r^2\cos^2(\theta)+v^2)^{-\frac{\alpha}{2}}}\diff v}{\diff \theta}\nnb\\
& \hspace{1.5cm}\exp\left(-2\lambda_l\int_{0}^{r} 1-{ e^{-2\mu_b\sqrt{r^2-u^2}}}e^{-2\mu_b\int_{\sqrt{r^2-u^2}}^{\infty}\frac{Tr^\alpha(u^2+v^2)^{-\frac{\alpha}{2}}}{1+{Tr^\alpha (u^2+v^2)^{-\frac{\alpha}{2}}}}\diff v}\diff u\right)\nnb\\
&\hspace{1.5cm}\exp\left(-2\lambda_l\int_{r}^{\infty}1-e^{-2\mu_b\int_{0}^{\infty}\frac{Tr^\alpha(u^2+v^2)^{-\frac{\alpha}{2}}}{1+{Tr^\alpha (u^2+v^2)^{-\frac{\alpha}{2}}}}\diff v}\diff u\right)\diff r.\label{part2}
\end{align}
	The above expression gives (b) of Eq. \eqref{L9:1}.
\par As a result, adding \eqref{31''} and \eqref{part2} gives the probability
that the typical vehicular user is associated with a vehicular base station
and is covered under the Palm distribution of $ \Psi_u $.
\end{IEEEproof}

\begin{figure}
	\centering
	\includegraphics[width=0.5\linewidth]{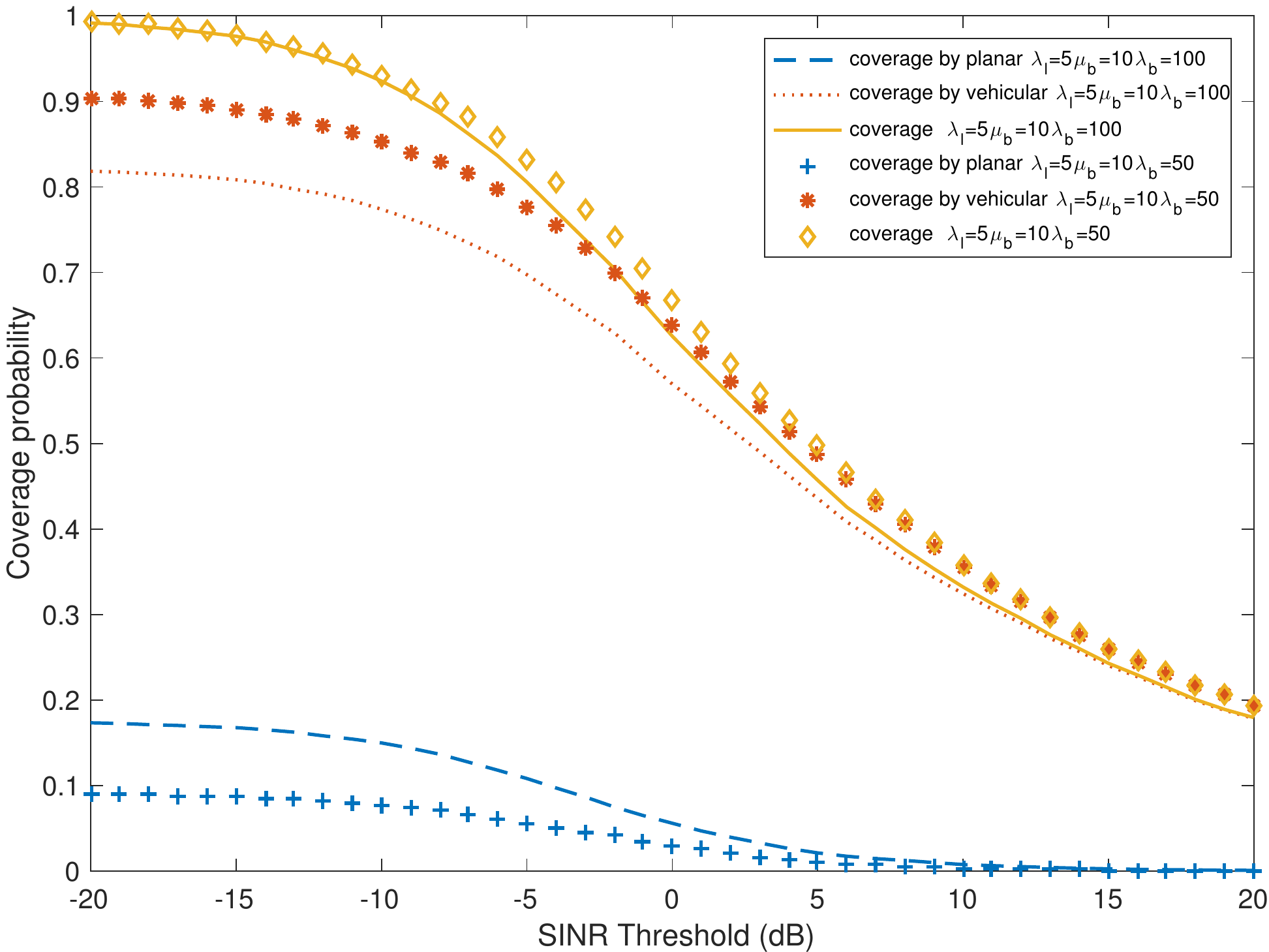}
	\caption{Coverage probability of the typical user under the Palm distribution of $ \Psi_u. $}
	\label{fig:figure10}
\end{figure}

Fig. \ref{fig:figure10} illustrates the coverage probability of the typical user under the Palm distribution of $ \Psi_u.$
Even when the spatial densities of planar base stations and vehicular base stations are the same,
the coverage probability provided by planar base stations is much smaller than the one provided by vehicular base stations.
We can interpret this as follows. In general, under the Palm distribution of $ \Psi_u$, most vehicular users
are associated with vehicular base stations and only a small fraction of vehicular users are associated with
planar base stations (Proposition 1). When the vehicular users are associated with planar base stations,
they are easily located at the {cell boundaries} where interference from nearby vehicular base stations
is significant. Note that even with twice more planar base stations, $\lambda_l\mu_b=50 $ and $ \lambda_b=100, $
the spatial average of the SINR provided by planar base stations is smaller than that provided by vehicular base stations.
This phenomenon is related to the topology of lines and it should have an impact on the architecture of future cellular networks. 

\section{Coverage: Palm Probability of $ \Phi_u+\Psi_u $}\label{S:5}
Here, we summarize Lemmas 1 to 4 to evaluate the coverage probability 
of the typical user under the Palm distribution of $ \Phi_u+\Psi_u. $
\subsection{Coverage Probability of the Typical User: Deconditioning}
\begin{theorem}
	The coverage probability of the typical user is given by
	\begin{align}
	p_c&=\frac{\lambda_u}{\lambda_u+\lambda_l\mu_u }\uta{\bP_{\Phi_u}^0(\SINR>T,X_\star\in\Phi_b)}+\frac{\lambda_u}{\lambda_u+\lambda_l\mu_u }\utb{\bP_{\Phi_u}^0(\SINR>T,X_\star\in\Psi_b)}\nnb\\
	&+ \frac{\lambda_l\mu_u}{\lambda_u+\lambda_l\mu_u }	\utc{\bP_{\Psi_u}^0(\SINR>T,X_\star\in\Phi_b)}+ \frac{\lambda_l\mu_u}{\lambda_u+\lambda_l\mu_u }	\utd{\bP_{\Psi_u}^0(\SINR>T,X_\star\in\Psi_b)},\nnb
	\end{align}	
	where (a), (b), (c), and (d) are given by  \eqref{L5:1},  \eqref{L6:1},   \eqref{L8:1}, and \eqref{L9:1}, respectively.
\end{theorem}
\begin{IEEEproof}
	The proposed model is jointly stationary. Using \eqref{jointpalm}, the coverage probability is 
	\begin{align}
	 p_c&=\frac{\lambda_u}{\lambda_u+\lambda_l\mu_u }\bP_{\Phi_u}^0(\SINR>T) 	+ \frac{\lambda_l\mu_u}{\lambda_u+\lambda_l\mu_u }\bP_{\Psi_u}^0(\SINR>T)\nnb.
	\end{align}	
	As a result, combining Lemmas from 1 to 4 gives the proof of the theorem.
\end{IEEEproof}
\subsection{Interpretation of the Palm Coverage Probabilities}
\begin{figure}[htb]
	\centering
	\begin{tabular}{@{}cc@{}}
		\includegraphics[width=.48\textwidth]{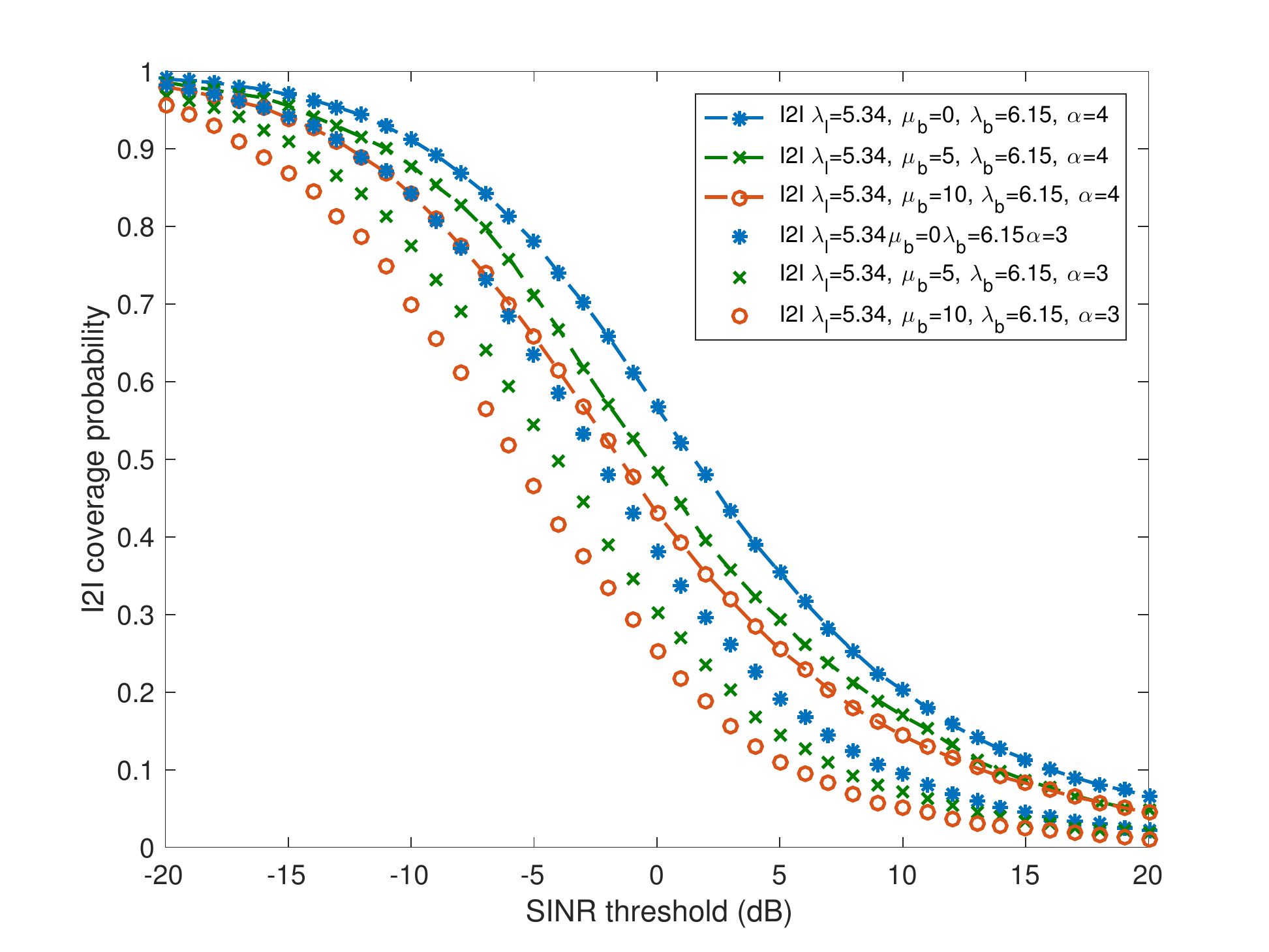} &
		\includegraphics[width=.48\textwidth]{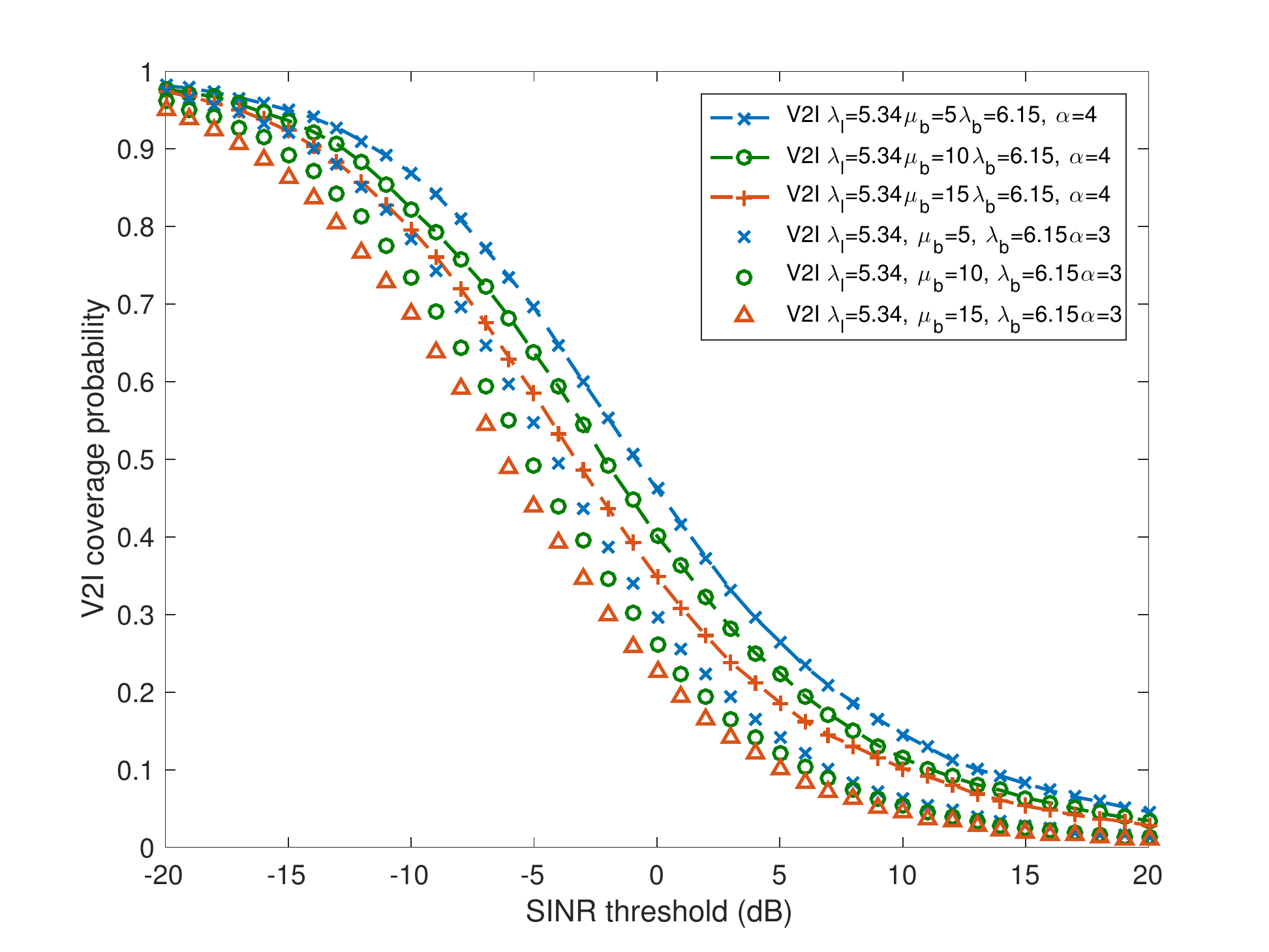} \\
		\includegraphics[width=.48\textwidth]{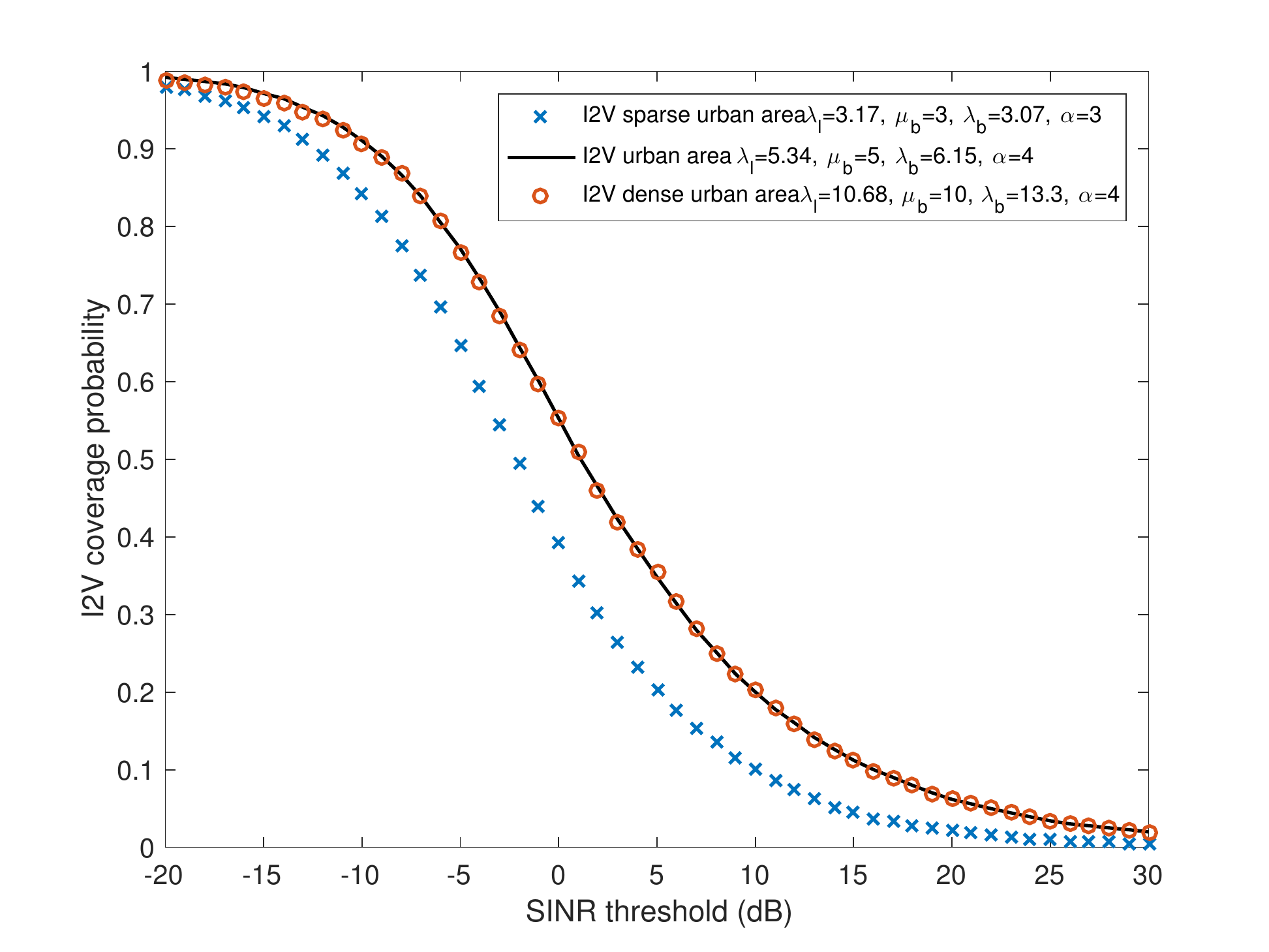} &
		\includegraphics[width=.48\textwidth]{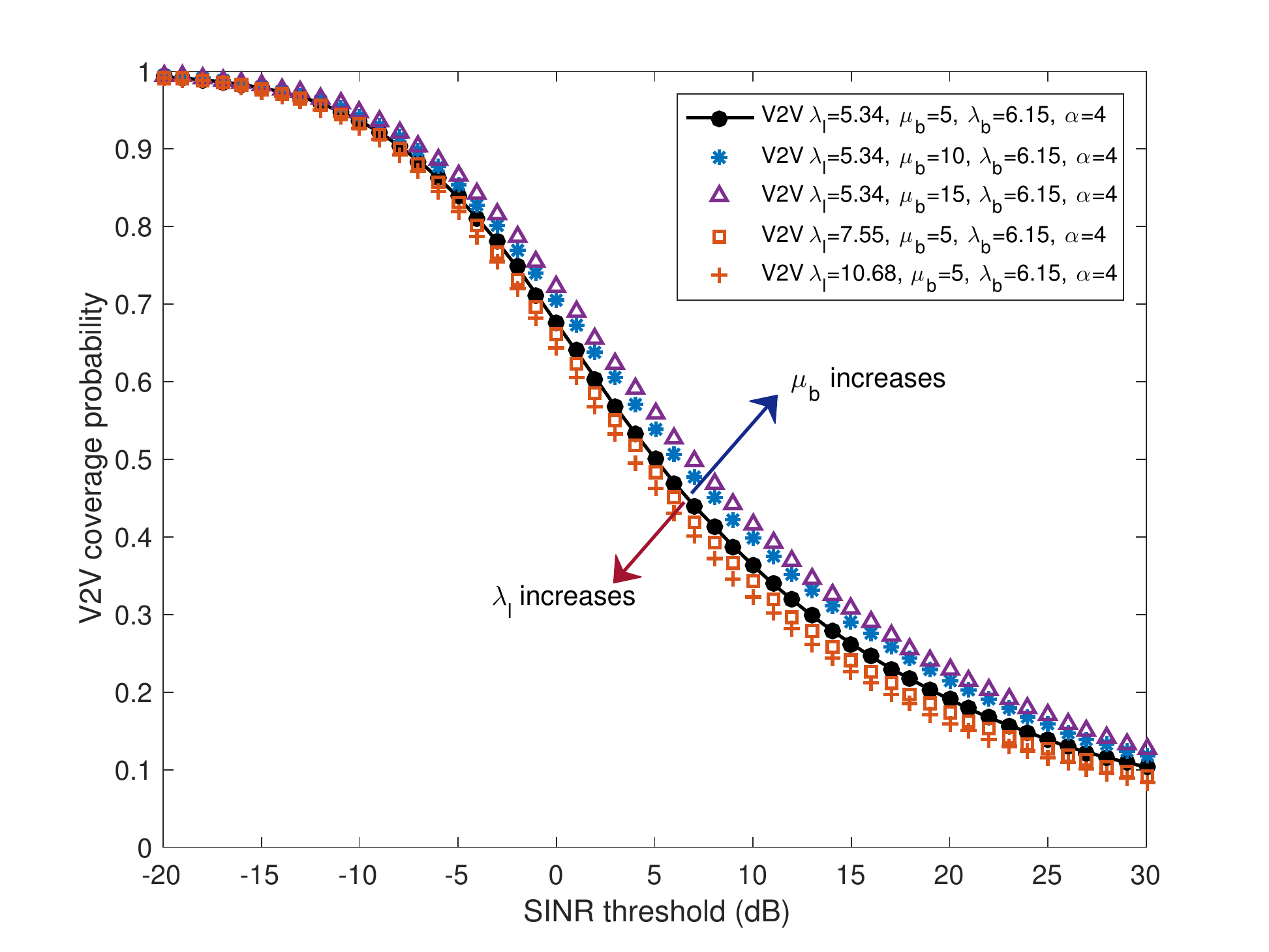} 
	\end{tabular}
	\caption{Illustration of the coverage probabilities of all typical links presented in the proposed model.}\label{fig:7}
\end{figure}

In a vehicular cellular network, downlink transmissions
are divided into four groups: vehicle-to-vehicle, vehicle-to-infrastructure, infrastructure-to-vehicle,
and infrastructure-to-infrastructure. Each group has different service requirements and operation protocols.
For instance, the vehicle-to-vehicle communication requires higher reliability for a shorter distance
than the other groups do because it might be used in safety applications \cite{karagiannis2011vehicular}.
In the course of this paper, the coverage expression of each communication group was actually derived\footnote{Here, we consider the Poisson distributed planar users as infrastructure users by noting their topological equivalence.}.
\begin{itemize}
	\item The coverage probability of link from vehicle base station-to-vehicle user (V2V) is 
	 \begin{equation}\label{38}
	 \bP_{\Psi_u}^0(\SINR>T|X_\star\in\Psi_b)=\frac{ \bP_{\Psi_u}^0(\SINR>T,X_\star\in\Psi_b)}{\bP_{\Psi_u}^0(X_\star\in\Psi_b)}.
	 \end{equation}
	\item The coverage probability of link from infrastructure base stations-to-vehicle users (I2V) is 
\begin{equation}
\bP_{\Psi_u}^0(\SINR>T|X_\star\in\Phi_b)=\frac{ \bP_{\Psi_u}^0(\SINR>T,X_\star\in\Phi_b)}{\bP_{\Psi_u}^0(X_\star\in\Phi_b)}.
\end{equation}

 	\item The coverage probability of link from vehicle base station-to-infrastructure user (V2I) is 
 	 \begin{equation}
 	 \bP_{\Phi_u}^0(\SINR>T|X_\star\in\Psi_b)=\frac{ \bP_{\Phi_u}^0(\SINR>T,X_\star\in\Psi_b)}{\bP_{\Phi_u}^0(X_\star\in\Psi_b)}.
 	 \end{equation}
 	
	\item The coverage probability of link from infrastructure base station-to-infrastructure user (I2I) is
	 \begin{equation}\label{42}
	 \bP_{\Phi_u}^0(\SINR>T|X_\star\in\Phi_b)=\frac{ \bP_{\Phi_u}^0(\SINR>T,X_\star\in\Phi_b)}{\bP_{\Phi_u}^0(X_\star\in\Phi_b)}.
	 \end{equation}
	
\end{itemize}
All numerators and denominators in Eq. \eqref{38}-\eqref{42} are already provided in previous sections.
\par  Fig. \ref{fig:7} illustrates the coverage probability of all possible links present in the proposed network. Let us first investigate the coverage of infrastructure users. The upper left figure illustrates the I2I coverage probability. As the number of vehicular base stations increases, $ \mu_b=5,10,15/\text{km}  $, the typical infrastructure user experiences a lower coverage probability. Similarly, as illustrated in the upper right figure, the V2I coverage probability diminishes as the density of vehicular base stations grows.
\par Then, let us investigate the coverage of vehicular users. In the lower right figure, we observe that as the number of vehicular base station increases the V2V coverage probability increases. This is because the typical vehicle is more likely to have an increased desired signal power. In contrast, as the size of road blocks shrinks, $ \lambda_l=5.34,7.55, 10.88/\text{km} $, the V2V coverage probability decreases. This happens because the typical vehicular user is associated with vehicular base stations on the same line at the same average  distance; nevertheless, it is exposed to an increased interference from vehicular base stations on the other lines. Notice that these trends are sensitive to network parameters and path loss model. For instance, when $ \mu_b $ is very small, the trend explained above could be reversed; with a high probability, the typical vehicle might be associated with vehicular base stations on {the other lines} since there might be no close vehicular base station on its own line. In the lower left figure, we consider sparse, normal, and dense scenarios where the density of road blocks and planar base stations vary. This figure shows that the I2V coverage probability is higher in  dense areas than in sparse areas.


\section{Conclusion}\label{S:7}
This paper provides a representation of heterogeneous cellular networks consisting of vehicular base stations,
vehicular users, planar base stations, and planar users which takes their topological characteristics into account.
Vehicular base stations and users on roads are modeled by Cox point processes given by independent Poisson point
processes conditionally on a Poisson line process; on the other hand, planar base stations and users are
modeled by independent Poisson point processes. We characterized the network performance seen by 
typical users of both kinds by deriving the association probability and coverage probability under
the Palm distribution of each user point process. This allows us to obtain analytical expressions for the coverage probabilities of all possible links present in the proposed network architecture.
 \par This brings a first understanding on the basic performance metrics such as user association probability,
interference power distribution, and coverage probability, but also provides a framework for future
research on practical scenarios in future cellular architecture based on vehicles.
For instance, this framework could be used to address resource allocation problems in
the presence of simultaneous vehicular and planar links with possibly different quality of service requirements.
 
\section*{Acknowledgment}
{\small{This work is supported in part by the National Science Foundation under Grant
No. NSF-CCF-1218338 and an award from the Simons Foundation (\#197982), both
to the University of Texas at Austin.}}


\bibliographystyle{IEEEtran}
\bibliography{ref_FBedit}
\appendices
\section{Alternate Proof of Lemma 2}\label{A:1}

In this proof, Palm calculus is used to derive the coverage probability.
Define 
\[ I_{\Phi}^{r}(x)=Tr^\alpha\sum_{X_i\in\Phi\setminus B(x,r)}H_i\|X_i-x\|^{-\alpha}. \]
This is the interference created by a point process $ \Phi $ seen by a point $ x $,
with a protected region of radius $ r $.
Then, 
\begin{align}
p&=\bE_{\Phi_u}^0\left(\ind_{\SINR>T}\right)\nnb\\
&=\bE_{\Phi_u}^0\left(\sum_{X_i\in\Phi_u}\bP\left(H_i>T\|X_i\|^\alpha I_{\Phi_b+\Psi_b}^{\|X_i\|}(0)\right)\right)\nnb\\
&\ea\bE\left[\sum_{X_i\in\Phi_b} e^{-I_{\Phi_b+\Psi_b}^{\|x\|}(0)}1_{\Phi_b(B(0,\|X_i\|))=0}1_{\Psi_b(B(0,\|X_i\|))=0}\right]\nnb\\
&\eb\lambda_l\mu_b\int_{\bR^2}\bE_{\Psi_b}^0\left[e^{-I_{\Phi_b}^{\|x\|}(-x)}e^{-I_{\Psi_b}^{\|x\|}(-x)}\ind_{\Phi_b(B(-x,\|x\|))=0}\ind_{\Psi_b(B(-x,\|x\|))=0}\right]\diff x\nnb\\
&\ec\lambda_l\mu_b\int_{\bR^2}\bE_{\Psi_b}^0\left[e^{-I_{\Psi_b}^{\|x\|}(-x)}\ind_{\Psi_b(B(-x,\|x\|))=0}\right]\bE_{\Phi_b}\left[e^{-I_{\Phi_b}^{\|x\|}(-x)}\ind_{\Phi_b(B(-x,\|x\|))=0}\right]\diff x\nnb\\
&\ed\lambda_l\mu_b\int_{\bR^2}\bE_{\Phi_b}\left[e^{-I_{\Phi_b}^{\|x\|}(-x)}\ind_{\Phi_b(B(-x,\|x\|))=0}\right]\bE_{\Psi_b}^{!0}\left[e^{-I_{\Psi_b}^{\|x\|}(-x)}\ind_{\Psi_b^{!0}(B(-x,\|x\|))=0}\right]\nnb\\
&\hspace{2.1cm}\bE_{\phi}\left[e^{-I_{\phi}^{\|x\|}(-x)}\ind_{\phi(B(-x,\|x\|))=0}\right]\diff x,\label{3333}
\end{align}
where we have (a) from independence and (b) from the Slivnyak-Little-Mecke-Mattes formula \cite{Baccelli2013Elements},
which holds for any stationary point process $ \Phi: $
\begin{align}
\bE\left[\int_{\bR^2}f(x,\theta_x\omega)\Phi(\diff x)\right]=\lambda_{\Phi}\int_{\bR^2} \bE^{0}\left[f(x,\omega)\right]\diff x.\nnb
\end{align} We obtain (c) from the independence between $ \Psi_b $ and $ \Phi_b $, and (d) from the fact that under Palm distribution of $ \Psi_b $, there exists a line passing through the origin. 
The first and second terms of Eq. \eqref{3333} are derived by following steps similar to those in the proof of Lemma 1 and are 
\begin{align}
\bE_{\Phi_b}\left[e^{-I_{\Phi_b}^{\|x\|}(-x)}\ind_{\Phi_b(B(-x,\|x\|))=0}\right]=&e^{-\pi\lambda_b\|x\|^2}e^{-2\pi\lambda_b\int_{\|x\|}^{\infty}\frac{T\|x\|^\alpha u^{1-\alpha}}{1+T\|x\|^\alpha u^{-\alpha}}\diff u},\label{12}\\
\bE_{\Psi_b}^{!0}\left[e^{-I_{\Psi_b}^{\|x\|}(-x)}\ind_{\Psi_b^{!0}(B(-x,\|x\|))=0}\right]=	&e^{-2\lambda_l\int_{0}^{\|x\|} 1-e^{-2\mu\sqrt{\|x\|^2-u^2}-2\mu_b\int_{\sqrt{r^2-u^2}}^\infty \frac{T\|x\|^\alpha{}\left(u^2+v^2\right)^{-\frac{\alpha}{2}} }{1+T\|x\|^\alpha\left(u^2+v^2\right)^{-\frac{\alpha}{2}} }\diff v}\diff u}\nnnl
&e^{-2\lambda_l\int_{\|x\|}^\infty 1-e^{-2\mu_b\int_{0}^\infty \frac{T\|x\|^\alpha{}\left(u^2+v^2\right)^{-\frac{\alpha}{2}} }{1+T\|x\|^\alpha{}\left(u^2+v^2\right)^{-\frac{\alpha}{2}} }\diff v}\diff u},\label{13}
\end{align}
respectively. The last term of \eqref{3333} is the Laplace transform of the
interference from the line that contains the origin.
Notice that the angle of the line process $ \phi $ is denoted by $ \theta $
and it is uniformly distributed  between 0 and $ \pi. $we have 
\begin{align}
&\bE_{\phi}\left[e^{-I_{\phi(\theta)}^{\|x\|}(-x)}\ind_{\phi(\theta)(B(-x,\|x\|)=0)}\right]\nnb\\
&\ee\int_{0}^{\pi} \bE\left[e^{-I_{\phi(\theta)}^{\|x\|}(-x)}\ind_{\phi(\theta)(B(-x,\|x\|)=0)}|\theta\right]f_{\Theta}(\theta)\diff \theta\nnb\\
&\ef\frac{2}{\pi}\int_{0}^{\pi/2}\bE_{\phi(\theta)}\left[\prod_{T_j\in\phi(\theta)\setminus B(-x,\|x\|)}e^{-H\|T_j+x\|^{-\frac{\alpha}{2}}}\right]\bP(\phi(\theta)(B(-x,\|x\|)=0))\diff \theta\nnb\\
&\eg\frac{2}{\pi}\int_0^{\pi/2}\bE_{\phi(\theta)}\left[\prod_{T_j\in\phi(\theta)\setminus B(-x,\|x\|)}e^{-H\|T_j+x\|^{-\frac{\alpha}{2}}}\right]e^{-2\mu_b\|x\|\sin(\theta)}\diff \theta\nnb\\
&\eh\frac{2}{\pi}\int_{0}^{\pi/2}e^{-2\mu_b\int_{\|x\|\sin(\theta)}^{\infty}\frac{T\|x\|^{\alpha}{(\|x\|^2\cos^2(\theta)+t^2)}^{-\frac{\alpha}{2}}}{1+T\|x\|^{\alpha}{(\|x\|^2\cos^2(\theta)+t^2)}^{-\frac{\alpha}{2}}}\diff t}e^{-2\mu_b\|x\|\sin(\theta)}\diff \theta,\label{234}
\end{align}
where we have (e) from the conditioning on the angle of $ \phi, $,
(f) from Slivnyak's theorem, and (g) from the property of the Poisson point process with intensity $ \mu. $
For (g), we utilize the fact that, without loss of generality, we can consider $ -x $
is on the $ Y$ axis. Then, the length of arc created by the ball centered at $ -x $ and 
the line $ l_{0,\theta} $ is equal to $ 2\|x\|\sin(\theta) $, where $ \theta $ is
the angle between the line and the positive $ X $ axis.
To derive (h), we use the Laplace transform of the Poisson point process
with line intensity $ \mu_b $. Finally, we incorporate the derived 
Eqs. \eqref{12}, \eqref{13}, and \eqref{234}, and integrate \eqref{3333}
with respect to the polar coordinate to complete the proof.

\end{document}